\documentclass[aps,pra,twocolumn,amsmath,amssymb,showpacs,floatfix,superscriptaddress]{revtex4-1}
\usepackage{times}

\usepackage[utf8x]{inputenc}
\usepackage[english]{babel}
\usepackage[T1]{fontenc}
\usepackage{lmodern}
\usepackage{amssymb}
\usepackage{bbold}
\usepackage{amsmath}
\usepackage{amsthm}
\usepackage[pdftex]{graphicx}
\usepackage{epstopdf}
\usepackage{braket}

\usepackage{dcolumn}
\usepackage{bm}
\usepackage{color}
\usepackage{relsize,dsfont,mathrsfs,empheq,verbatim,upgreek}
\usepackage{etoolbox}
\usepackage[caption = false]{subfig}

\usepackage{pgfplots}

\newcommand{\Rv}{\boldsymbol{R}}

\newcommand{\mathd}{\mathrm{d}}
\newcommand{\Tr}{\mathrm{Tr}}

\makeatletter
\renewcommand{\fnum@figure}{Fig. \thefigure}
\makeatother

\begin{document}

\title{Entropy Production and the Role of Correlations in Quantum Brownian Motion}

\author{Alessandra Colla}

\affiliation{Institute of Physics, University of Freiburg, 
Hermann-Herder-Stra{\ss}e 3, D-79104 Freiburg, Germany}

\author{Heinz-Peter Breuer}

\affiliation{Institute of Physics, University of Freiburg, 
Hermann-Herder-Stra{\ss}e 3, D-79104 Freiburg, Germany}

\affiliation{EUCOR Centre for Quantum Science and Quantum Computing,
University of Freiburg, Hermann-Herder-Stra{\ss}e 3, D-79104 Freiburg, Germany}

\begin{abstract}
We perform a study on quantum entropy production, different kinds of correlations, and their interplay
in the driven Caldeira-Leggett model of quantum Brownian motion. The model, taken with a large
but finite number of bath modes, is exactly solvable, and the assumption of a Gaussian initial state 
leads to an efficient numerical simulation of all desired observables in a wide
range of model parameters. Our study is composed of three main parts. We first compare two popular 
definitions of entropy production, namely the standard weak-coupling formulation originally proposed 
by Spohn and later on extended to the driven case by Deffner and Lutz, and the always-positive expression
introduced by Esposito, Lindenberg and van den Broeck, which relies on the knowledge of the evolution of the bath.  
As a second study, we explore the decomposition of the Esposito et al. entropy production into 
system-environment and intra-environment correlations for different ranges of couplings and temperatures. Lastly, 
we examine the evolution of quantum correlations between the system and the environment, measuring
entanglement through logarithmic negativity.
\end{abstract}
 
\date{\today}

\maketitle

\section{Introduction}\label{sec:Intro}

The analysis of open quantum systems which are coupled to an environment plays a key role in many
applications of quantum mechanics \cite{Breuer2002}. One of the central goals of the theory is the development of an
efficient description of the reduced dynamics of the open system in which the degrees of freedom of the
environment have been eliminated by performing a partial trace over the environmental Hilbert space. 
A typical result is an effective equation of motion for the reduced density matrix representing the quantum state 
of the open system, e.g. a Markovian or non-Markovian quantum master equation
\cite{Rivas2014a,Breuer2016a,deVega2017}.

While highly efficient in many cases of interest, such a treatment of open quantum systems completely relies on the 
degrees of freedom of the open system and does not allow any access to the environmental degrees of freedom.
However, there can of course be physically relevant quantities which require knowledge about the total 
system-environment state and/or the reduced environmental state. Indeed, such quantities arise, for example, 
in the construction of certain expressions for the entropy production in open quantum systems coupled to heat
baths, or in the study of the role of correlations generated by the system-environment interaction.
To tackle those questions a possible strategy is to analyze paradigmatic model systems. Here, we employ the
Caldeira-Leggett model \cite{Caldeira1983} of quantum Brownian motion modelling a central harmonic oscillator, 
representing the open system, which is coupled to a reservoir of harmonic oscillators describing the environment.
In addition, we also examine the influence of a driving force acting on the central oscillator.
This is a well-known integrable model which has been studied extensively in the literature \cite{Grabert1988}.
Recently, we have used this model to carry out a detailed study of non-Markovianity in quantum Brownian
motion \cite{Einsiedler2020}. Taking a large but finite number of environmental harmonic oscillator modes, the 
Caldeira-Leggett model can be solved exactly by a transformation to normal modes \cite{Ullersma1966} which, 
together with the assumption of Gaussian initial states, leads to an efficient method for the evaluation of 
general physical quantities of the total system in a wide range of model parameters such as temperature, 
system-environment coupling, driving frequency and amplitude.

In the present paper we investigate three main topics. First, we compare different definitions for 
quantum entropy production arising in the quantification of the degree of irreversibility of quantum processes
and in the formulation of the second law of quantum thermodynamics \cite{Binder2018}.
Namely, we compare the original
expression proposed by Spohn \cite{Spohn1978,Spohn_Lebowitz1978} and the later generalization to the 
driven case by Deffner and Lutz (DL entropy production) \cite{Lutz2011} with the entropy production suggested by 
Esposito, Lindenberg and van den Broeck (ELB entropy production) \cite{Esposito2010}. 
A similar comparison for the Caldeira-Leggett model without driving and without frequency renormalization 
has been carried out in Ref.~\cite{Pucci2013}. Quite interestingly, we find that the two definitions, which
converge in the high-temperature and weak-coupling limit in the undriven case, are incompatible
when driving is present on the central oscillator.

The second topic is the decomposition of the ELB entropy production into three parts discussed recently
\cite{Esposito2019}, representing the mutual information between system and environment, the mutual information
describing the intra-environmental correlations, and the sum of the distances of the individual bath modes from their
initial values (measured in terms of relative entropy). The present study demonstrates that, by contrast to the
findings reported in \cite{Esposito2019}, in our system the intra-environmental correlations need not provide
the dominant contribution to the entropy production. In particular, driving the central oscillator leads to a
drastic change of the relative size of the various contributions since the mutual information is not affected by driving, 
while the relative entropy quantifying the shift of the bath modes is influenced by driving.
Finally, our third topic is the entanglement between the central
oscillator and the environmental modes which is generated by the system-environment interaction and
quantified by the logarithmic negativity \cite{Vidal2002}.

The manuscript is organized as follows. In Sec.~\ref{sec:TheModel} we briefly discuss the model system and the strategy used
to determine all desired observables. In Sec.~\ref{sec:Eprod} we review different definitions for the entropy 
production proposed in the literature, the decomposition of the ELB entropy production, and the quantification
of system-environment correlations in terms of mutual information and logarithmic negativity.
Our numerical simulation results are presented and discussed in detail in Sec.~\ref{sec:results}. Finally, we summarize the
results and draw our conclusions in Sec.~\ref{sec:conclu}.

\section{Model system}\label{sec:TheModel}

\subsection{Microscopic Hamiltonian}

The microscopic description of quantum Brownian motion is provided by the Caldeira-Leggett model, which represents the dissipative dynamics of a single particle coupled linearly to a bath of harmonic oscillators. We focus here on the integrable case of a quadratic potential for the central particle, which then becomes a harmonic oscillator itself. Without restriction we can set all masses equal to one. Furthermore, we take the bath to be of a finite size corresponding to $N$ modes, such that the Hamiltonian reads \cite{Caldeira1983}
\begin{eqnarray}\label{HCL}
H=&\underbrace{\frac{1}{2}p_0^2+\frac{1}{2}\omega_0^2 x_0^2}_{\text{$H_S$}} 
+\underbrace{\sum_{n=1}^N \left(\frac{1}{2}p_n^2+\frac{1}{2}\omega_n^2 x_n^2 \right)}_{\text{$H_E$}} \\ \nonumber
&\underbrace{-x_0\sum_{n=1}^N \kappa_n x_n}_{\text{$H_I$}}+\underbrace{x_0^2 \sum_{n=1}^N 
\frac{\kappa_n^2}{2\omega_n^2}}_{\text{$V_c$}} \;, 
\end{eqnarray}
where we have added $V_c$ as the counter term accounting for the renormalization of the central oscillator frequency due to the interaction 
with the bath \cite{Grabert1988}. As a consequence, we call $\omega_0$ the renormalized frequency. We can then formally absorb the counter term into the frequency of the central oscillator by defining the bare frequency:
\begin{equation}\label{renfreq}
\omega_b = \sqrt{\omega_0^2 + \sum_{n=1}^N 
\frac{\kappa_n^2}{\omega_n^2}} \; .
\end{equation}
The influence of the bath on the dynamics of the central oscillator is modeled by the spectral density
\begin{equation}
J(\omega)=\sum_{n=1}^N \frac{\kappa_n^2}{2 \omega_n}\delta(\omega-\omega_n) \;,
\end{equation}
which contains information both on the density of oscillators in the bath at a certain frequency and on the strength of the coupling between the central system and such oscillators. We set this discretized quantity to reproduce, in the continuum and infinite bath limit, an Ohmic spectral density with a sharp cutoff
\begin{equation}
\label{spectral}
J(\omega)=\frac{2\gamma}{\pi}\omega \Theta(\Omega-\omega) \;,
\end{equation}
where $\gamma$ is then the coupling constant and $\Omega$ is the high-frequency cutoff. In the limit of infinite cutoff, $\gamma$ also takes the role of the damping constant for the central oscillator. We define the finite bath frequencies by sampling them uniformly from zero to a large maximal frequency $ \omega_{\mathrm{max}}$, like done in \cite{Pucci2013}, such that $\omega_n= n \Delta$ for $n \in \{1,...,N\}$, where $\Delta= \omega_{\mathrm{max}}/N$. To reproduce the coupling induced by the spectral density, we say that, for a number of modes $N$ large enough,
\begin{equation}
\int_{\omega_n-\Delta/2}^{\omega_n+\Delta/2} J(\omega) \mathrm{d} \omega =  \frac{\kappa_n^2}{2 \omega_n} \approx \Delta J(\omega_n) \; ,
\end{equation}
such that we can define each coupling to be
\begin{equation}
\kappa_n = \sqrt{2\Delta \omega_n J(\omega_n)} \; .
\end{equation}
The choice of a finite size environment naturally entails the appearance of Poincaré recurrences, which then occur at a time scale $t_{\mathrm{max}} = 2 \pi N /  \omega_{\mathrm{max}}$, de facto limiting our time availability for a reliable study, and requiring us to choose suitably large values of $N$.

In the course of the paper, we at times consider the addition of an external driving force. While the event of such force having influence on the bath certainly has some physical relevance and is worthy of exploration \cite{Grabert2018}, we choose to address the (still relevant) case of the driving force $F(t)$ acting on the central oscillator only, so that the Hamiltonian \eqref{HCL} is modified in the following:
\begin{equation}
\label{HCLdr}
H(t)=\frac{1}{2}p_0^2+\frac{1}{2}\omega_b^2 x_0^2 - F(t)x_0 + H_E + H_I \; . 
\end{equation}
In particular, we consider the case of finite-time driving, with an enveloped sinusoidal driving force $F(t)$ of the form 
\begin{eqnarray}
F(t)= \begin{cases}
F_0 \sin(\omega_f t + \phi) \sin^2(\Omega_f t) \;, \: \: \: t \leq {\pi \over \Omega_f} \;, \\
0 \;, \;  \: \: \:t > {\pi \over \Omega_f} \;,
\end{cases} 
\end{eqnarray}
where $F_0$ is the pulse height, $\pi/ \Omega_f$ the pulse duration and $\omega_f$ the driving frequency.

\subsection{Exact evolution}\label{sec:evolution}

\subsubsection{Solution of the model}
The Caldeira-Leggett model \eqref{HCL} is exactly solvable for a finite bath of dimension $N$ by means of a transformation to normal modes. This was first suggested by Ullersma in \cite{Ullersma1966} and further discussed in \cite{Pucci2013} for almost exactly the same model as we take here, the only difference being the absence of the counter term $V_c$. We employ the same prescription, only switching the physical frequency $\omega_0$ with the frequency \eqref{renfreq} balancing renormalization. The core idea is the following: by defining the vector of position and momentum operators $\Rv^{\mathrm{T}}=(x_0, x_1, ... , x_N, p_0, p_1, ... , p_N)$, one can represent the Hamiltonian \eqref{HCL} with the help of a $2(N+1)\times 2(N+1)$ matrix $\mathcal{H}$, like so
\begin{equation}
H= {1\over 2} \Rv^{\mathrm{T}} \mathcal{H} \Rv  \;,
\end{equation}
where
\begin{equation}
\mathcal{H} = \begin{bmatrix}
\mathcal{H}_x & \mathbb{0}_{n+1} \\
 \mathbb{0}_{n+1}  &  \mathbb{1}_{n+1} 
\end{bmatrix}  \;,
\end{equation}

\begin{equation}\label{Hmat}
\mathcal{H}_x = \begin{bmatrix}
\omega_b^2 & - k_1 & - k_2 & ... &  -k_N \\
- k_1 & \omega_1^2 & 0 &  & 0 \\
- k_2 & 0 & \omega_2^2 &  & 0 \\
\vdots & & & \ddots& \vdots \\
- k_N & 0 & 0 & ... & \omega_N^2
\end{bmatrix}  \; .
\end{equation}
Then there exists an orthogonal, symplectic matrix $S \in \mathrm{Sp}(2(N+1), \mathbb{R})\cap \mathrm{O}(2(N+1))$ such that $D=S^{\mathrm{T}}\mathcal{H}S$ is diagonal: in fact, $\mathcal{H}_x$ is symmetric, thus can be diagonalized by an orthogonal matrix $Z$ such that $D_x=Z^{\mathrm{T}}\mathcal{H}_xZ$ is diagonal; one can then define $S= Z \oplus Z$, which is orthogonal and symplectic and diagonalizes $\mathcal{H}$. The Hamiltonian can then be written as
\begin{equation}
H= {1\over 2} \Rv'^{\mathrm{T}} S^{\mathrm{T}}\mathcal{H}S \Rv'  \;,
\end{equation}
such that $S$ induces a transformation on $\Rv$:
\begin{equation}
\Rv'= S^{\mathrm{T}} \Rv = Z^T\oplus Z^T \Rv \; ,
\end{equation}
which in turn define transformed position and momentum operators $\{({x'}_{\mu}, {p'}_{\mu})\}_{\mu=0}^N$, so that the system in this basis is a collection of decoupled harmonic oscillators:
\begin{equation}
H=\sum_{\mu=0}^N \frac{1}{2}{p'_{\mu}}^2+\frac{1}{2}z_{\mu}^2 {x'_{\mu}}^2 \;, 
\end{equation}
where $\{z^2_{\mu}\}_{\mu=0}^N$ are the diagonal entries of $D_x$, i.e. the eigenvalues of $\mathcal{H}_x$. One can then easily solve the Heisenberg equations of motion in the new basis,
\begin{eqnarray}\label{Heisx}
\dot{x}'_{\mu}(t) &=& p'_{\mu}(t) \\ \label{Heisp}
\dot{p}'_{\mu}(t) &=& - z_{\mu}^2 x'_{\mu}(t)
\end{eqnarray}
to obtain
\begin{eqnarray}
x'_{\mu}(t) &= x'_{\mu}(0)\cos(z_{\mu} t) + {p'_{\mu}(0)\over z_{\mu}}\sin(z_{\mu} t) \\
p'_{\mu}(t) &= p'_{\mu}(0)\cos(z_{\mu} t) - {x'_{\mu}(0) z_{\mu}}\sin(z_{\mu} t)
\end{eqnarray}
and then transform back into the old operators $\{({x}_{\mu}, {p}_{\mu})\}_{\mu=0}^N$, so that the exact solution for the system reads:
\begin{eqnarray}\label{xevol}
x_{\mu}(t) &= \sum_{\rho}\left[ \dot{A}_{\mu \rho}(t) x_{\rho}(0) + {A}_{\mu \rho}(t) p_{\rho}(0)\right] \\ \label{pevol}
p_{\mu}(t) &= \sum_{\rho}\left[ \ddot{A}_{\mu \rho}(t) x_{\rho}(0) + \dot{A}_{\mu \rho}(t) p_{\rho}(0) \right] \; ,
\end{eqnarray}
where
\begin{eqnarray}\label{A}
A_{\mu \rho}(t) &= \sum_{\nu} Z_{\mu \nu} Z_{\rho \nu} {\sin{z_{\nu} t } \over z_{\nu}} \; .
\end{eqnarray}

Introducing driving only slightly modifies the approach to the solution. The driving term in the Hamiltonian can in fact be added as
\begin{equation}
- F(t) x_0 = - \boldsymbol{F}(t) \cdot \Rv \; ,
\end{equation}
with $\boldsymbol{F}(t) = (F(t), 0 , ... , 0)^{\mathrm{T}}$; then, the symplectic transformation induces a scrambling in the force vector $ \boldsymbol{F}'(t) = S^{\mathrm{T}}  \boldsymbol{F}(t)$,  so that in the new coordinates one is left with a system of decoupled harmonic oscillators which are this time also driven:
\begin{equation}
H=\sum_{\mu=0}^N \left( \frac{1}{2}{p'}_{\mu}^2+\frac{1}{2}z_{\mu}^2 {x'}_n^2 - F'_{\mu}(t)x'_{\mu} \right) \;.
\end{equation}
Now the solution for each driven harmonic oscillator reads
\begin{eqnarray}\label{xevold}
x_{\mu}(t) &= \sum_{\rho} \left[ \dot{A}_{\mu \rho}(t) x_{\rho}(0) + {A}_{\mu \rho}(t) p_{\rho}(0)\right] + I_{\mu}(t) \\ \label{pevold}
p_{\mu}(t) &= \sum_{\rho} \left[ \ddot{A}_{\mu \rho}(t) x_{\rho}(0) + \dot{A}_{\mu \rho}(t) p_{\rho}(0)\right] + \dot{I}_{\mu}(t) \; ,
\end{eqnarray}
where $A_{\mu \rho}(t)$ is given by \eqref{A}, and
\begin{eqnarray}
{I}_{\mu}(t) &= \sum_{\nu} Z_{\mu \nu} Z_{0 \nu} \int_0^t {\sin{z_{\nu} (t-s) } \over z_{\nu}} F(s) \mathrm{d}s \; .
\end{eqnarray}
It is clear how the above procedure can readily be extended to treat a driven bath, which is however not the subject of this work.

\subsubsection{Initial conditions and Gaussian states}

Gaussian states are particularly useful when studying the exact evolution of systems modelled by quadratic Hamiltonians, since an initial Gaussianity of the state is preserved by the evolution, thus simplifying the description of the state at all times. For the study of the Caldeira-Leggett model, we therefore make use of an initial Gaussian state, such that its evolution is described at all times by its first moments $\braket{x_{\mu}}$, $\braket{p_{\mu}}$ and its second moments, i.e. the covariance matrix $\sigma$, which we define in our notation as:
\begin{equation}
\sigma_{\mu \nu}^{(\xi \eta)}= {1\over 2} \braket{\{ \xi_{\mu}, \eta_{\nu} \}} - \braket{\xi_{\mu}}\braket{\eta_{\nu}} \; ,
\end{equation}
where $\xi , \eta = x , p$.
We consider, as initial conditions, an uncorrelated Gaussian state:
\begin{equation}\label{rho0}
\rho_{SE}(0)= \rho_S(0)\otimes\rho_{E}^{\mathrm{eq}} \; ,
\end{equation}
where $\rho_{E}^{\mathrm{eq}}= \mathrm{e}^{-\beta H_E}/Z_E$, with $\beta = 1/ \mathrm{k_B} T$, is the bath equilibrium Gibbs state at temperature $T$. In terms of first and second moments, this corresponds to the following relations $\forall n, m \in \{1,...,N\}$:
\begin{eqnarray}\nonumber
&\braket{x_n(0)} = \braket{p_n(0)} = 0   \; , \; \; \sigma_{nm}^{(xp)}(0)= 0 \\\label{initialB} 
& \sigma_{nm}^{(xx)}(0)= {1 \over 2 \omega_n} \coth{\left( \omega_n \over 2 \mathrm{k_B} T\right)} \delta_{nm}  \; ,\\ \nonumber
& \sigma_{nm}^{(pp)}(0)= {\omega_n \over 2 } \coth{\left( \omega_n \over 2 \mathrm{k_B} T\right)} \delta_{nm} \; .
\end{eqnarray}

While the initial absence of system-bath correlations and the state of thermal equilibrium of the bath are required conditions for the computation of entropy production (see Sec.~\ref{sec:Eprod}), we have freedom in the choice of the central oscillator initial state $\rho_S(0)$. We choose the system to be initially in the ground state with respect to its physical frequency $\omega_0$, i.e.
\begin{eqnarray}\label{initialS}
&\braket{x_0(0)} = \braket{p_0(0)} = 0   \; , \; \; \sigma_{00}(0)= \begin{bmatrix}
{1\over 2\omega_0} & 0 \\
0 & {\omega_0\over 2}
\end{bmatrix} \; .
\end{eqnarray}
The evolution of the total state $\rho_{SE}(t)$ can then be completely described in terms of first and second moments with the help of equations \eqref{xevold} and \eqref{pevold}. For the means, this is simply:
\begin{align}\label{mxevold}
\braket{x_{\mu}(t)}&= \sum_{\rho}\left[ \dot{A}_{\mu \rho}(t) \braket{x_{\rho}(0)} + {A}_{\mu \rho}(t) \braket{p_{\rho}(0)} \right] + I_{\mu}(t) \\ \label{mpevold}
\braket{p_{\mu}(t)} &= \sum_{\rho} \left[ \ddot{A}_{\mu \rho}(t) \braket{x_{\rho}(0)} + \dot{A}_{\mu \rho}(t) \braket{p_{\rho}(0)} \right] + \dot{I}_{\mu}(t) \; ,
\end{align}
so that, in general, the evolution of the first moments can depend on $A_{\mu 0}(t)$ and its time derivatives. With our choice of the initial reduced state \eqref{initialS}, though, each oscillator sees a displacement in time that is only due to the effect of the driving force, from which the term $I_{\mu}(t)$ originates. For the covariance matrix, on the contrary, the driving contribution has no effect at all. This is a consequence of the linearity of the system and of the restriction to Gaussian states, as already noticed in \cite{Einsiedler2020} for the limit of infinite bath modes. The evolution of the covariance matrix is then of the following form:
\begin{eqnarray}\nonumber
\sigma_{\mu \nu}^{(\xi\eta)}(t) &=& \sum_{\rho \sigma}\Big[ 
\dot{B}_{\mu \rho}^{(\xi\eta)}(t) \dot{C}_{\nu \sigma}^{(\xi\eta)}(t) \sigma_{\rho \sigma}^{(xx)}(0) \\ \nonumber
&&+
\left( 
	\dot{B}_{\mu \rho}^{(\xi\eta)}(t) C_{\nu \sigma}^{(\xi\eta)}(t) + B_{\mu \sigma}^{(\xi\eta)}(t) \dot{C}_{\nu \rho}^{(\xi\eta)}(t) 
	\right) \sigma_{\rho \sigma}^{(xp)}(0) \\	
&&+ B_{\mu \rho}^{(\xi\eta)}(t) C_{\nu \sigma}^{(\xi\eta)}(t) \sigma_{\rho \sigma}^{(pp)}(0)
\Big] \; , 
\end{eqnarray}
where $B$ and $C$ can represent either the matrix \eqref{A} or its time-derivative, specifically
\begin{eqnarray}\nonumber
&B^{(xx)}= A(t) \; , \; \;
C^{(xx)}= A(t) \\
&B^{(xp)}= A(t)  \; , \; \;
C^{(xp)}= \dot{A}(t)\\ \nonumber
&B^{(pp)}= \dot{A}(t)  \; , \; \;
C^{(pp)}= \dot{A}(t) \; .
\end{eqnarray}

\section{Entropy production and correlations}\label{sec:Eprod}

\subsection{Different definitions of entropy production}

There is no general consensus on how to define proper thermodynamic quantities at a quantum level, especially in the more general framework where the coupling between the system and the environment is allowed to be strong. Different definitions of work and heat consequently lead to different forms of entropy production, which is canonically defined as:
\begin{equation}
\Delta_i S(t) = \Delta S_S(t) - {1\over \mathrm{k_B} T} \delta Q_S(t) \; ,
\end{equation}
where $S_S(t)$ is the von Neumann entropy associated to the system, $T$ is the temperature of the bath and $\delta Q_S$ is the heat exchange. What we refer to here as the ``standard approach'' \cite{Landi2020}, which is the one typically used in the context of weak coupling, sees the heat exchange as:
\begin{equation}
\delta Q_S^{\mathrm{st}}(t) := \int_0^t \mathrm{d} s \Tr \{ H_S(s) \dot{\rho}_S(s) \} \; .
\end{equation}
If the Hamiltonian $H_S$ of the system is taken to be time independent, the heat exchange is then identical to the change in internal energy of the system, i.e. $\Delta U_S(t)= \braket{H_S}_{\rho_S(t)} - \braket{H_S}_{\rho_S(0)}$. This gives rise to the well-known form of the entropy production originally proposed by Spohn \cite{Spohn1978}:
\begin{equation}\label{Spohn}
\Delta_i S^{\mathrm{Sp}}(t) = S(\rho_S(0)|| \rho_S^{\mathrm{eq}}) - S(\rho_S(t)|| \rho_S^{\mathrm{eq}})\; ,
\end{equation}
where $S(\rho||\sigma)$ is the relative entropy and $\rho_S^{\mathrm{eq}}$ is the Gibbs state associated to the system Hamiltonian $H_S$. This expression has then been extended to the case of time dependent Hamiltonians by Deffner and Lutz \cite{Lutz2011}, leading to the following expression:
\begin{eqnarray}\nonumber
\Delta_i S^{\mathrm{DL}}(t) &=  S(\rho_S(0)|| \rho_{S}^{\mathrm{eq}}(0)) - S(\rho_S(t)||\rho_S^{\mathrm{eq}}(t)) \\ \label{DLutz}
&-\int_0^t \mathd s \Tr \{ \rho_S(s)\partial_s \ln \rho_S^{\mathrm{eq}}(s)\}  \; ,
\end{eqnarray}
where we defined 
\begin{equation}
\rho_S^{\mathrm{eq}}(t)= e^{- \beta H_S(t) }/Z_S(t)
\end{equation} 
as the instantaneous Gibbs state associated to the Hamiltonian at time $t$. This standard approach is extensively in use in the weak coupling regime, although it is currently seen as a set-back by many that the entropy production in this framework can drastically reach negative values once the strong coupling regime is entered. A different proposal for entropy production which is instead always positive has been developed by Esposito, Lindenberg and van den Broeck \cite{Esposito2010}, where the heat exchange is then defined as:
\begin{eqnarray}\label{HeatEsp}
\delta Q_S^{\mathrm{ELB}}(t) :=  - \Delta U_E(t) 
=\braket{H_E}_{\rho_E(0)} -\braket{H_E}_{\rho_E(t)}\; ,
\end{eqnarray}
which, in the assumption of a time-independent bath Hamiltonian, corresponds to taking into account in the heat exchange also the contribution coming from the interaction:
\begin{equation}
\delta Q_S^{\mathrm{ELB}}(t) =  \int_0^t \mathrm{d} s \Tr \{ [H_S(s) + H_I(s)] \dot{\rho}_S(s) \}\; .
\end{equation}
Taking the heat exchange definition \eqref{HeatEsp} and imposing an uncorrelated initial state with the bath in thermal equilibrium, $\rho_{SE}(0)= \rho_S(0)\otimes \rho_E^{\mathrm{eq}}$, leads to an expression for entropy production which is positive at all times (although it can oscillate):
\begin{equation}\label{Esposito}
\Delta_i S^{\mathrm{ELB}}(t) = S(\rho_{SE}(t)|| \rho_S(t)\otimes \rho_E^{\mathrm{eq}}) \; .
\end{equation}
From the origin of the two definitions of entropy production, it is clear how the difference between them should vanish in the limit of some coupling strength going to zero. While expression \eqref{Esposito} has the practical disadvantage of depending on the time evolution of the bath -- which makes it hard to be computed for the vast majority of model systems considered -- it is by some considered an extension of the standard approach, and a more accurate definition for entropy production, also in view of its natural interpretation as a quantum Landauer's principle \cite{Reeb2014}. 

A comparison between the two different approaches to entropy production has been already performed for the Caldeira-Leggett model in \cite{Pucci2013}, confirming their compatibility in the small coupling regime, for the case of an undriven central oscillator. One of the questions considered in this paper (see Sec.~\ref{sec:results}) is whether this compatibility still holds when driving is added.

\subsection{Contributions of entropy production}\label{sec:contrib}

Recently, the expression \eqref{Esposito} was more closely studied in \cite{Esposito2019} for a quantum dot coupled to fermionic baths, shedding some light on which quantities can significantly contribute to entropy production. It is there shown how the entropy production can be split into three main contributions:
\begin{eqnarray}\label{contributions}
\Delta_iS^{\mathrm{ELB}}(t)= I_{SE}(t) + I_{\mathrm{env}}(t) + D_{\mathrm{env}}(t) \; ,
\end{eqnarray}
where $I_{SE}$ corresponds to the mutual information between the system and the environment,
\begin{eqnarray}\nonumber \label{Ise}
I_{SE}(t) &=& S_S(t)+ S_E(t) - S_{SE}(t) \\
&=& \Delta S_S(t)+\Delta S_E(t) \; ,
\end{eqnarray}
and the terms $D_{\mathrm{env}}$ and  $I_{\mathrm{env}}$ make up the contribution due to the distance of the environment from its initial state:
\begin{eqnarray}\label{IenvDenv}
I_{\mathrm{env}}(t) + D_{\mathrm{env}}(t) = S(\rho_E(t)||\rho_E(0))\; .
\end{eqnarray}
What is argued in \cite{Esposito2019}, where the states of the central system considered (a two-level system) live in a finite-dimensional Hilbert space, is that the main contribution to the whole entropy production cannot consistently come from the mutual information between the system and the environment. 
In fact, the latter is strongly bounded from above by the inequality
\begin{eqnarray}\label{ArakiLieb}
I_{SE} \leq 2 \min \{S_S, S_E\} \; ,
\end{eqnarray}
which follows immediately from the Araki-Lieb inequality \cite{Nielsen2000} $S_{SE} \geq | S_S - S_E |$ and which, for large baths, implies
\begin{eqnarray}
I_{SE} \leq 2 \ln N \; ,
\end{eqnarray}
with $N$ the dimension of the Hilbert space of the system; on the contrary, entropy production itself can be time-extensive in certain systems. So the main part of entropy production should be given by the other two contributions in Eq. \eqref{IenvDenv}, which are defined as
\begin{eqnarray}\label{Ienv}
I_{\mathrm{env}}(t) = \sum_n S_{E_n}(t) - S_E(t)\; ,
\end{eqnarray}
i.e. the mutual information among the bath modes, describing the intra-environment correlations, and
\begin{eqnarray}\label{Denv}
D_{\mathrm{env}}(t) =  \sum_n S(\rho_{E_n}(t)||\rho_{E_n}(0)) \; ,
\end{eqnarray}
which represents the sum of the distances of the individual bath modes from their initial state. For the system considered in \cite{Esposito2019}, it was numerically found that, of these two terms, the environment mutual information $I_{\mathrm{env}}$ is the determining contribution. 

We thought it interesting to perform an analogue analysis on the contribution to the entropy production for the Caldeira-Leggett model, even though the argument based on the Araki-Lieb inequality \eqref{ArakiLieb} is non applicable for this model due to the infinite dimension of the Hilbert space of the central oscillator. Nevertheless, it will be clear from the numerical simulations in section \ref{sec:results} that for some range of parameters the result is the same ($I_{\mathrm{env}}$ is the main contribution), while for others not. Moreover, the driven setting also makes a radical difference in this regard. In fact, since the mutual information quantities $I_{SE}$ and $I_{\mathrm{env}}$ cannot be affected by local operators (and indeed depend only on the covariance matrix), they will be left unchanged by the addition of driving, which only acts locally on first moments. The distance of the environment modes $D_{\mathrm{env}}$, on the other hand, is written in terms of relative entropies and is indeed affected by how the mean values of the environment are pushed away from their initial values by the action of the driving force. Especially for strong and resonant driving, $D_{\mathrm{env}}$ increases with respect to its non-driven value and has therefore the possibility to triumph over $I_{\mathrm{env}}$ as the dominant contribution of entropy production.

\subsection{Entropy production and Gaussian states}
The use of Gaussian states has the practical advantage of enabling easy computation of key quantities in entropy production. Much of this comes from the ability to go to normal modes with the help of Williamson's theorem. The von Neumann entropy for some $n$-mode Gaussian state, for example, can be computed by defining the symplectic eigenvalues $\{\nu_i\}_{i=1}^n$ of the $n$-mode covariance matrix $\sigma$, which is a positive definite $2n \times 2n$ matrix, and thus can be brought to diagonal form by a symplectic transformation:
\begin{equation}
S^{\mathrm{T}} \sigma S = \Lambda \; ,
\end{equation}
where the diagonal matrix $\Lambda$ is the direct sum of the form
\begin{equation}
\Lambda = \Lambda_n \oplus \Lambda_n\; ,
\end{equation}
and the diagonal elements of $\Lambda_n$ are defined as the symplectic eigenvalues $\{\nu_i\}_{i=1}^n$. Operationally, the symplectic eigenvalues can be found by selecting the positive eigenvalues of the matrix $\mathrm{i} \Omega \sigma$ \cite{Adesso2007}, where $\Omega$ here represents the standard symplectic matrix. Then, the von Neumann entropy of the Gaussian state $\rho$ associated to the covariance matrix $\sigma$ can be written as:
\begin{align}\nonumber
S(\rho) =  \sum_{i=1}^n & \left[ \left(\nu_i +{ 1 \over 2}\right) \ln \left(\nu_i + {1 \over 2}\right) \right. \\
&- \left. \left(\nu_i - {1 \over 2}\right) \ln \left(\nu_i - {1 \over 2}\right) \right] \; .
\end{align}
This exact method is used, for example, for the calculation of the total bath entropy $S_E$. In the particular case of a one-mode Gaussian state -- as, for example, when evaluating the system entropy $S_S$ or the single bath mode entropies $S_{E_n}$ -- there is only one symplectic eigenvalue $\nu = \sqrt{\det \sigma}$. In those cases, the covariance matrix to consider is naturally the one obtained by selecting the entries of the total covariance matrix related to the specific mode considered. 

It is also useful for the evaluation of \eqref{Spohn}, \eqref{DLutz} and \eqref{Denv}, to know how to compute the relative entropy between two Gaussian states. This can be done generically by exploiting the Gibbs-exponential expression of the density matrix for a Gaussian state \cite{Banchi2015}:
\begin{equation}
\rho_{\alpha} = {1 \over Z_{\alpha} } \exp \left[ -{1 \over 2} (\Rv - \bar{\Rv}_{\alpha})^{\mathrm{T}} \mathcal{G}_{\alpha} (\Rv - \bar{\Rv}_{\alpha}) \right] \; , 
\end{equation}
where the entries of $\bar{\Rv}_{\alpha}$ are given by the first moments, while $\mathcal{G}_{\alpha}$ and $Z_{\alpha}$ depend on the covariance matrix, in particular
\begin{equation}\label{covG}
\mathcal{G}_{\alpha} = 2 \mathrm{i} \Omega \coth^{-1}(2 \mathrm{i} \sigma_{\alpha} \Omega) \; .
\end{equation}
Then it is easy to see that the relative entropy between two Gaussian states $\rho_1$ and $\rho_2$ is given by
\begin{eqnarray}\nonumber
S(\rho_1||\rho_2) =& S(\rho_2) - S(\rho_1) +{1\over 2} \braket{(\Rv - \bar{\Rv}_2)^{\mathrm{T}} \mathcal{G}_2(\Rv - \bar{\Rv}_2)}_{\rho_1} \\
&- {1\over 2} \braket{(\Rv - \bar{\Rv}_2)^{\mathrm{T}} \mathcal{G}_2(\Rv - \bar{\Rv}_2)}_{\rho_2} \; ,
\end{eqnarray}
so that everything is explicitly formulated in terms of means and covariance matrices. In our case, the relative entropy is always calculated for one-mode Gaussian states, which makes it easier to find the explicit expression for the matrix \eqref{covG}.

This setup takes care of the computation of most of the quantities mentioned. The rest, e.g. the mean value of the internal energy, can be calculated directly using the time evolution of single mean values of $x$ and $p$, and of the single entries of the covariance matrix.

\subsection{Entanglement and correlations with Gaussian states}\label{sec:entanglement}

As we have discussed, correlations happen to be an important feature in recent discussion on entropy production; a natural question to ask is then whether or not these correlations are of classical or quantum nature. Answering this question in a generic context constitutes in and of itself a whole active research area \cite{Adesso2016}. Successful separability criteria, for example, or a proper entanglement measure -- i.e., an entanglement monotone which can always detect its presence -- are tools that are hard to find in the vast majority of circumstances \cite{Horodecki2009}. While restricting the analysis of the topic to Gaussian states can somehow circumscribe the problems, the quantification of bipartite entanglement in continuous variable systems is also not yet completely resolved \cite{Adesso2007}.

Among others, a famous, necessary condition for separability of bipartite systems is the Peres-Horodecki criterion, also known as the Positive Partial Transpose (PPT) criterion \cite{Peres1996, Horodecki1997}. It consists of taking the partial transposition of the density matrix (with respect to one of the two systems in the bipartition) and looking at its eigenvalues: a separable state has a partial transpose which has all non-negative eigenvalues. In other words, the appearance of a negative eigenvalue in the partial transpose implies the presence of entanglement. As already noted,  this is a sufficient condition for entanglement, but, for most situations, not a necessary one. Cases in which it also happens to be necessary are limited, e.g. the $2\times 2$ and $2\times 3$ dimensional cases. 

For Gaussian states, the PPT criterion proves to be a quite powerful tool. In fact, the criterion happens to be also a sufficient condition for separability for states that pertain to systems of 1 vs $n$-modes \cite{Simon2000, Werner2001}. This works quite nicely for our application to the Caldeira-Leggett model, as we precisely deal with a central system of one mode coupled to a bath of $N$ modes. While quantum correlations in the bath are still not accessible, we are at least able to properly detect existing entanglement between the system and the reservoir. 

A PPT-related entanglement monotone which then works extremely well in our case is logarithmic negativity:
\begin{equation}\label{logneg}
E_{\mathcal{N}}(\rho) := \ln ||\tilde{\rho}||_1 \; ,
\end{equation}
where $\tilde{\rho}$ is the partially transposed density matrix, and $||\cdot||_1 $ is the trace norm. When the PPT criterion is equivalent to separability, like in our case, then $E_{\mathcal{N}}$ is a perfect measure for entanglement, as it is always different from zero when entanglement is present and does not increase under local operations and classical communication. Furthermore, it happens to be fairly easy to compute for Gaussian states. In fact, the PPT requirement in this framework amounts to asking the covariance matrix $\tilde{\sigma}$ of the partially transposed state to have all symplectic eigenvalues greater than one, $\tilde{\nu}_i \geq 1  \;\forall i$. For the case of Gaussian states of 1 vs $n$ modes, the logarithmic negativity \eqref{logneg} can then be computed as \cite{Adesso2007}:
\begin{equation}\label{logneg2}
E_{\mathcal{N}}(\rho) := \sum_{i=1}^{n+1} \max (0 , - \ln 2 \tilde{\nu}_i) \; .
\end{equation}

While logarithmic negativity is a good quantifier for entanglement, it does not allow us to derive a splitting of the total correlations between system and environment (i.e. mutual information $I_{SE}$) into quantum and classical correlations. This is, in general, a complex topic which is subject of extensive research, and which has seen the birth of many frameworks and definition for the quantification of classical and quantum correlations \cite{Modi2012, Bera2017}. Nonetheless, comparing the two measures ($I_{SE}$ and $E_{\mathcal{N}}$) and their trend with respect to the variation of some parameters (e.g. system-environment coupling or temperature) may still give hints about how entanglement influences the total system-bath correlations for the Caldeira-Leggett model. 

As mentioned, logarithmic negativity is not affected by local operations. Analogously to mutual information, therefore, the addition of driving -- which for Gaussian states has only a local displacement effect -- has no influence on the entanglement between the system and the reservoir. We therefore restrict our analysis on this subject to the non-driven case.

\section{Numerical results}\label{sec:results}

Taking into account a finite number of modes $N$ in the bath, it is possible to reproduce accurate results for all the different aforementioned quantities in the Caldeira-Leggett model, by following the procedure in Sec.~\ref{sec:evolution} and numerically diagonalizing the Hamiltonian matrix \eqref{Hmat} for different ranges of parameters. For this purpose, we use a large number of modes $N=400$ to better simulate the behaviour of an infinite bath. Moreover, the more modes we have at our disposal, the more time we have available before the appearance of recurrences; setting a maximal sampled frequency of $\omega_{\mathrm{max}}=40 \omega_0$ gives a time limit of $t_{\mathrm{max}} = 2 \pi N / \omega_{\mathrm{max}} \sim 63 / \omega_0$. To make sure that driving is present only in a finite amount of time such that relaxation is still achieved in the time remaining, we take the pulse duration $t_f$ to be half of the total time available, i.e. $t_f= \pi / \Omega_f = t_{\mathrm{max}}/2$. As for the cutoff $\Omega$, we choose to explore the case of large cutoff with respect to the renormalized frequency $\omega_0$. Using the sharp cutoff spectral density \eqref{spectral}, we take $\Omega$ to include all our sampled frequencies, i.e. $\Omega=\omega_{\mathrm{max}}=40 \omega_0$. The varying parameters in the following simulations are then coupling, temperature, driving amplitude and driving frequency $(\gamma, T, F_0, \omega_f)$. In these results, we have set $\omega_0=1$, so that all the quantities will be expressed in implicit units of the renormalized central frequency $\omega_0$.

\subsection{Two proposals for entropy production}\label{sec:res_proposals}

\begin{figure}[htp!]
  \includegraphics[clip,width=\columnwidth]{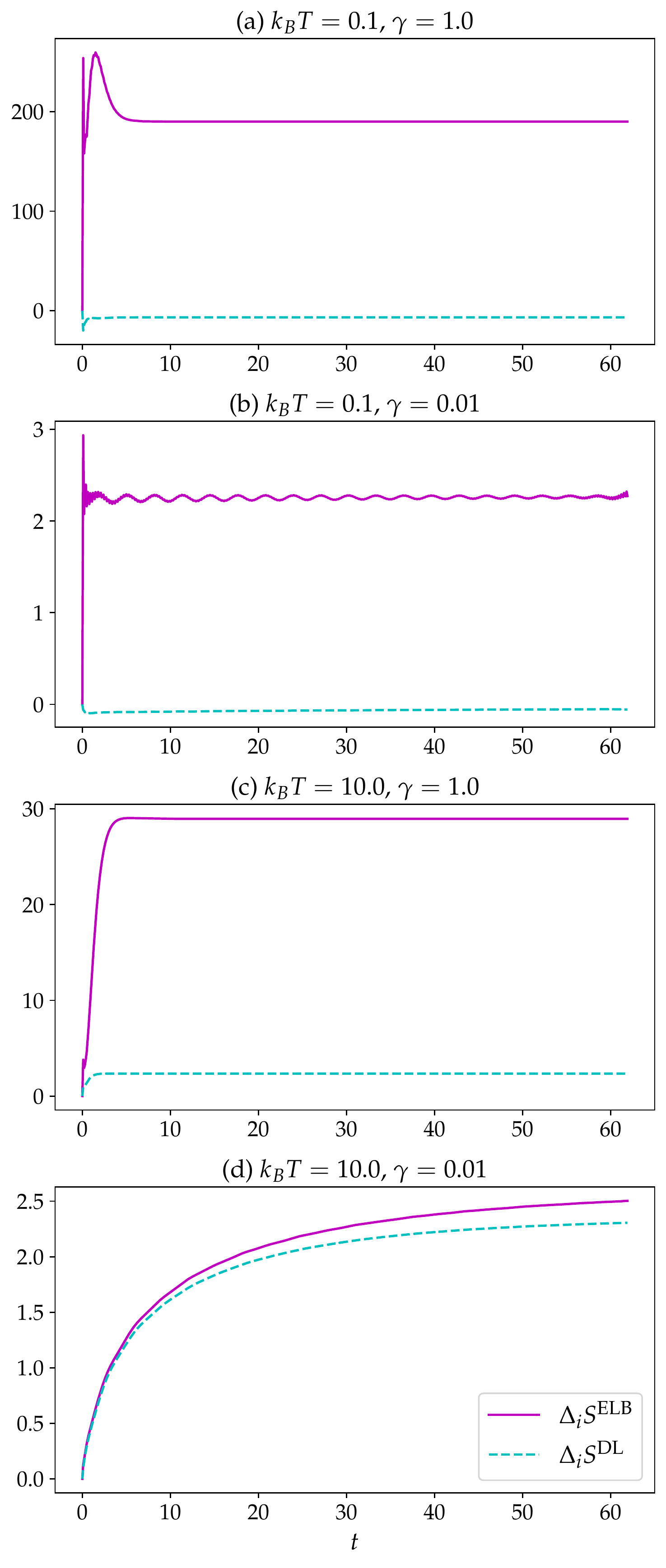}%
\caption{Comparison between the two definitions \eqref{DLutz} and \eqref{Esposito} for the Caldeira-Leggett model in absence of driving for four different ranges of coupling and temperature. The two quantities converge in the high temperature, weak coupling limit.}\label{fig:comp_nd}
\end{figure}

The two approaches for the formulation of thermodynamic quantities lead to the two different definitions \eqref{DLutz} and \eqref{Esposito} for entropy production. In the non-driven case, the two definitions, which can differ significantly, converge in the high temperature and weak coupling limit, as reported in \cite{Pucci2013}. In our framework and range of parameters, this can be seen in Figs. \ref{fig:comp_nd}a-\ref{fig:comp_nd}d. We recall that the difference between the two approaches is given by the inclusion (or not) of the contribution of the interaction Hamiltonian in the definition of heat exchange. The convergence is therefore intuitive by simply looking at the difference $\delta$ between them:
\begin{eqnarray}\nonumber
\delta (t) :&=& \Delta_i S^{\mathrm{ELB}}(t) - \Delta_i S^{\mathrm{DL}}(t)\\
&=& - {1\over \mathrm{k_B} T}  \int_0^t \mathrm{d} s \Tr \{ H_I(s) \dot{\rho}_{SE}(s) \} \; ,
\end{eqnarray}
which decreases with increasing temperature $T$ and decreasing interaction. In the model under study, the trend of $\delta$ with coupling strength and temperature can be seen explicitly by performing a perturbative expansion with respect to the coupling strength; we can extract the perturbation parameter $\lambda$ from the interaction, defined such that $\kappa_n \propto \lambda = \sqrt{\gamma}$, so that the interaction Hamiltonian for Caldeira-Leggett reads $\lambda H_I' + \lambda^2 V_c'$. The first term of the perturbation expansion is then of second order and reads $\delta^{(2)} (t) =  {\lambda^2 \over \mathrm{k_B} T}  \Delta(t)$, with
\begin{align}
\Delta(t) =  \mathrm{i} \int_0^t \mathd t_1 & \Tr \{ [\tilde{V}_c(t_1), \tilde{H}_0(t_1)] \rho_{SE}(0) \}    \\ \nonumber
 + \int_0^t  \mathd t_1 \int_0^{t_1} &\mathd t_2 \Tr \{ [\tilde{H}_I(t_2), [\tilde{H}_0(t_1),\tilde{H}_I(t_1)]] \rho_{SE}(0) \}  \; ,
\end{align}
where $\tilde{V}_c(t)$, $\tilde{H}_0(t)$, $\tilde{H}_I(t)$ denote respectively $V'_c$, $H_0(t)$ and $H'_I$ in the interaction picture with respect to $H_0(t):= H_S(t)+H_E$. Therefore
\begin{equation}\label{deltaprop}
\delta (t) \propto  {\gamma \over \mathrm{k_B} T}  \; ,
\end{equation}
such that $\delta \rightarrow 0$ for $\gamma, \beta \rightarrow 0$.

\begin{figure}[htp!]
  \includegraphics[clip,width=\columnwidth]{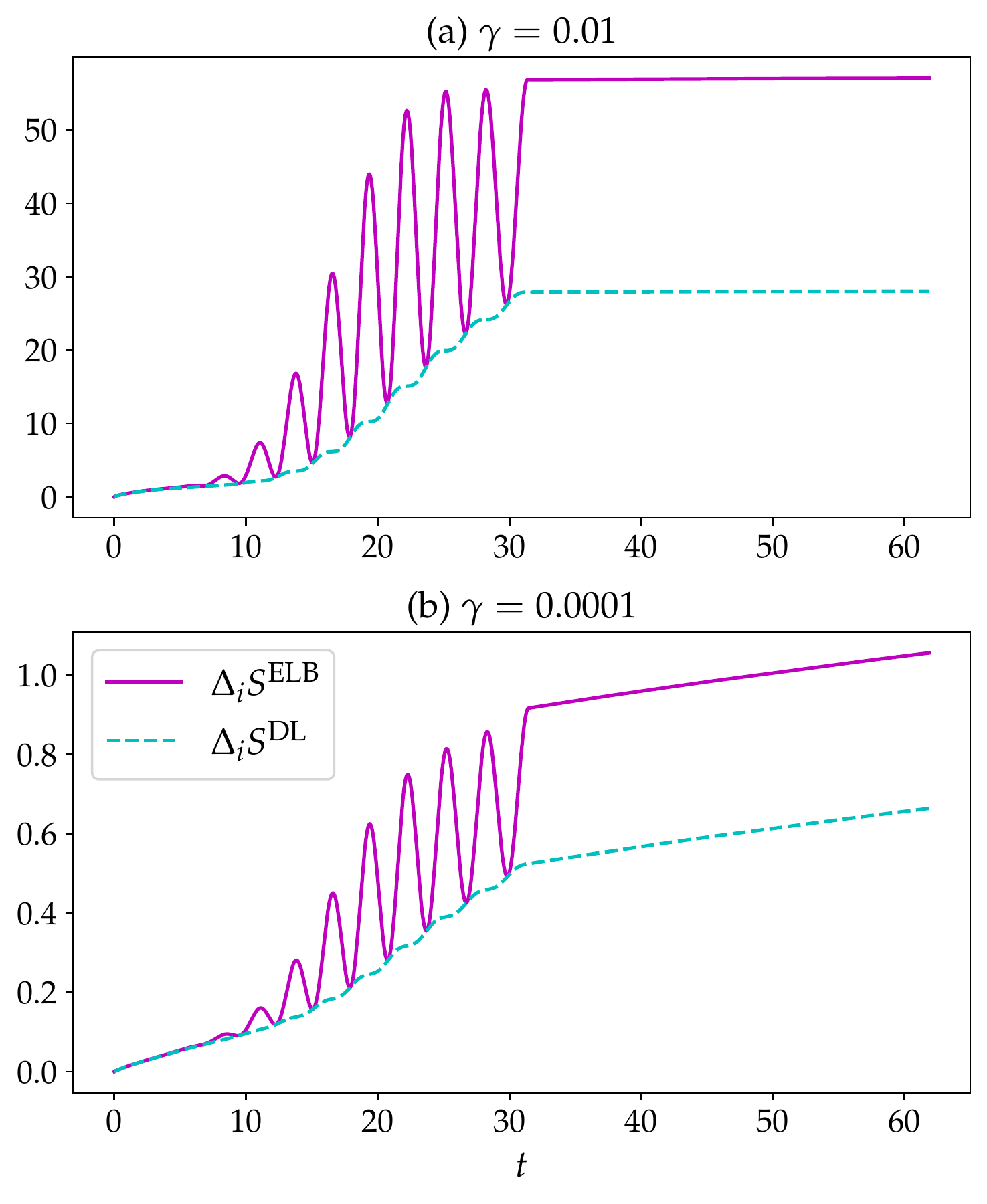}%
\caption{Comparison between the two definitions \eqref{DLutz} and \eqref{Esposito} under strong driving ($F_0=10$) in weak and ultra-weak coupling regimes, both at high temperature ($\mathrm{k_B}T/\omega_0=10$).}\label{fig:comp_dr}
\end{figure}

\begin{figure}[htp!]
 \includegraphics[clip,width=\columnwidth]{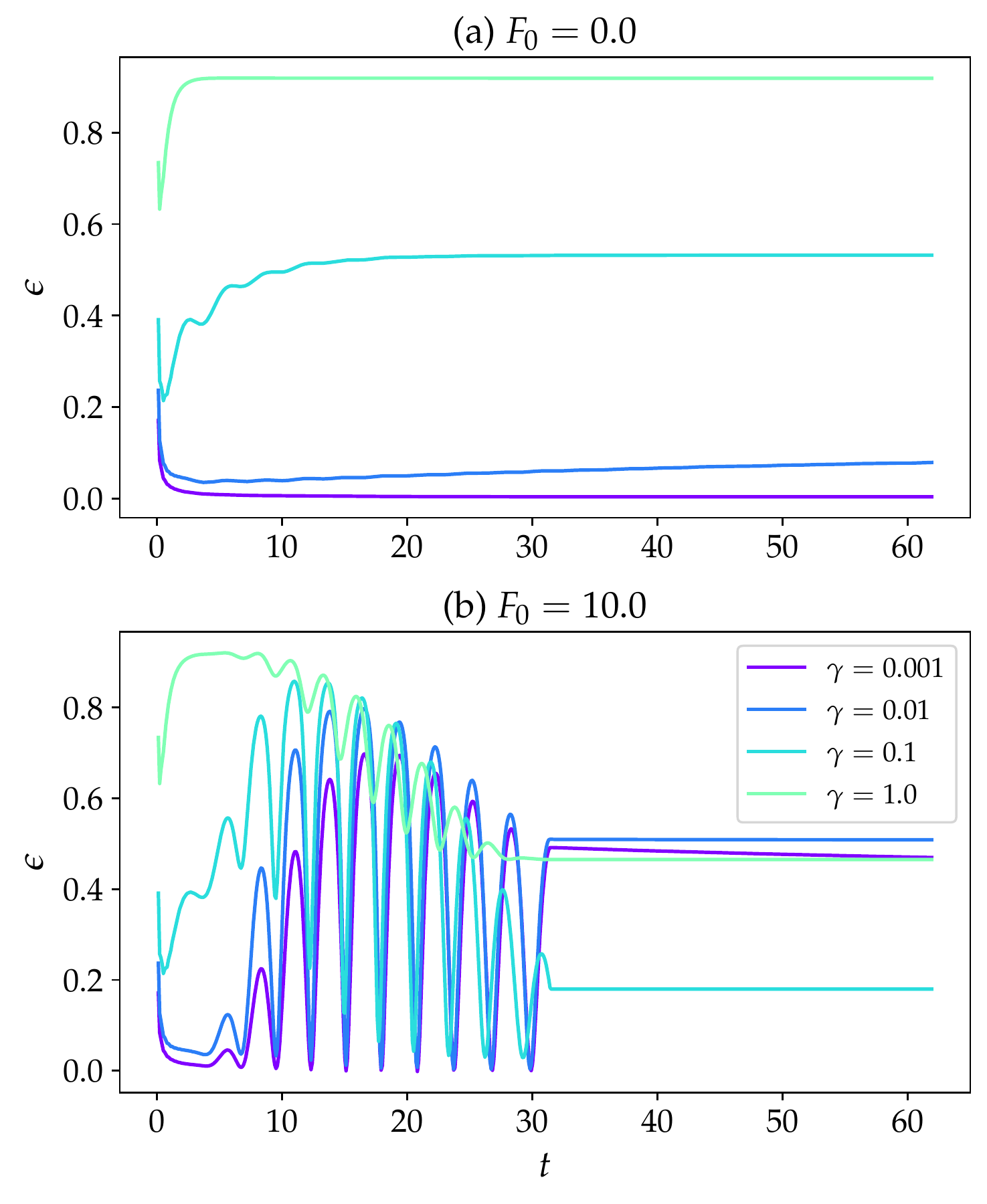}%
\caption{Time evolution of the relative error \eqref{relerr} between the two definitions of entropy production for different coupling parameters, at high temperature ($\mathrm{k_B}T/\omega_0=10$), in the driven and non-driven case.}\label{fig:rel_err}
\end{figure}

In the driven case, however, there seems to be a substantial difference. Let us focus on a fixed temperature $\mathrm{k_B}T/\omega_0=10$. The second definition $\Delta_i S^{\mathrm{ELB}}$ shows dramatic oscillations in the time interval in which the driving is switched on (we chose to directly look at strong driving, with an amplitude of $F_0=10$), while $\Delta_i S^{\mathrm{DL}}$ does not (see Fig. \ref{fig:comp_dr}a). This creates an oscillatory gap between the two definitions, which, surprisingly, does not disappear even in the ultra-weak coupling regime at high temperatures, as shown in Fig.~\ref{fig:comp_dr}b. 
While it is true, from the arguments above and from \eqref{deltaprop}, that the amplitude of the oscillatory gap tends to zero as the coupling constant $\gamma$ decreases, it also happens that the value of the entropy productions themselves drops to zero. One can further investigate this by looking at the evolution in time of the relative error between the two definitions
\begin{equation} \label{relerr}
\epsilon(t) := {\Delta_i S^{\mathrm{ELB}}(t) - \Delta_i S^{\mathrm{DL}}(t) \over \Delta_i S^{\mathrm{ELB}}(t)} \;. 
\end{equation}
While for the non-driven case the curve of $\epsilon$ flattens down to zero for decreasing values of $\gamma$ (Fig.~\ref{fig:rel_err}a), this does not happen when driving is added, as seen in Fig.~\ref{fig:rel_err}b, which shows non-zero oscillations even at extremely low coupling. This might be a symptom of a more serious underlying issue pertaining to the difference in the two definitions, and suggests that the two approaches might not simply be one the limiting case of the other. 

\subsection{Which quantity is the winner in entropy production?}\label{sec:res_contributions}

As explained in Sec.~\ref{sec:contrib}, we are interested in investigating the role of the constituents of the second definition for entropy production \eqref{Esposito} in the context of the Caldeira-Leggett model, namely which quantity between $I_{\mathrm{SE}}$ \eqref{Ise}, $I_{\mathrm{env}}$ \eqref{Ienv} and $D_{\mathrm{env}}$ \eqref{Denv} is the one mostly contributing to entropy production. For conciseness, we refer to \eqref{Esposito} simply as $\Delta_i S$ from now on.

\begin{figure}[htp!]
\includegraphics[clip,width=\columnwidth]{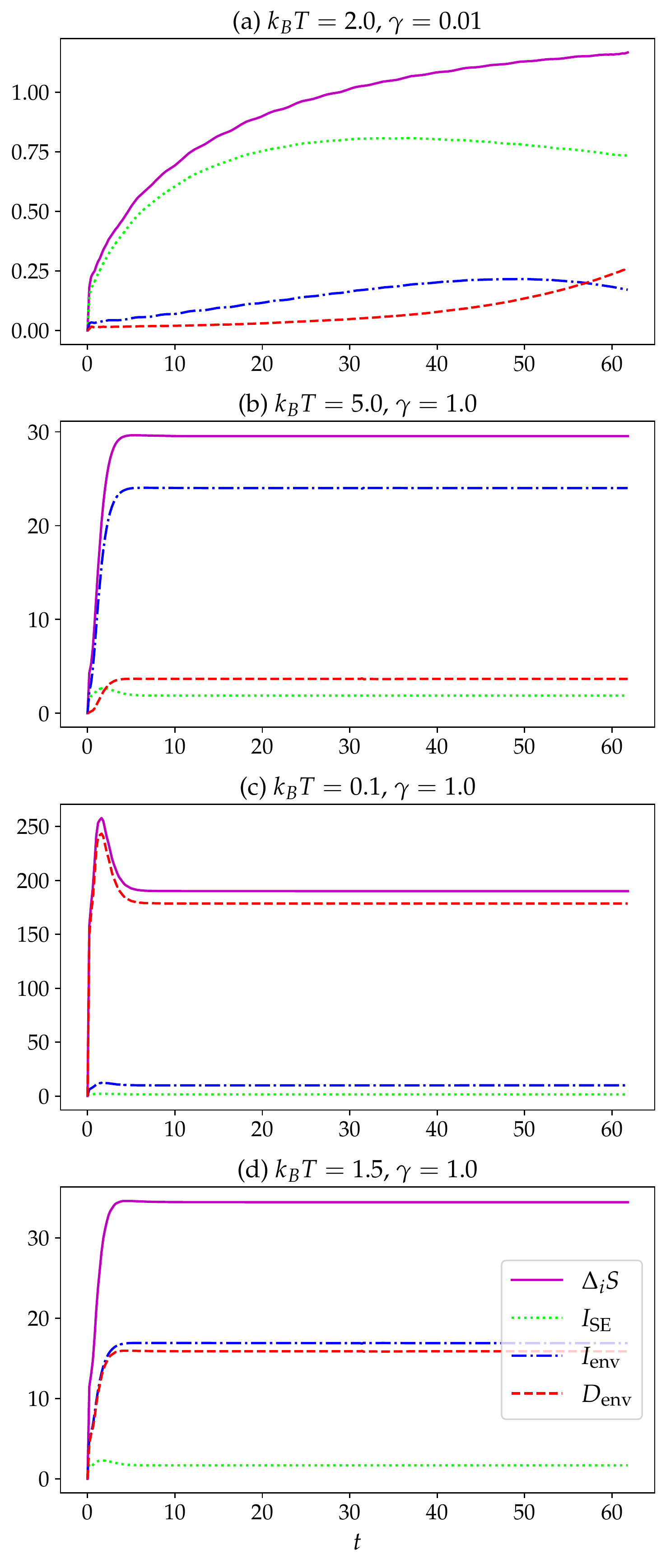}
\caption{Time evolution of the three contributions to entropy production $\Delta_iS^{\mathrm{ELB}}$ for different coupling parameters, for four different cases of coupling and temperature. Driving is here absent.}\label{fig:contr_nd}
\end{figure}

\begin{figure}
\includegraphics[width=0.9\linewidth]{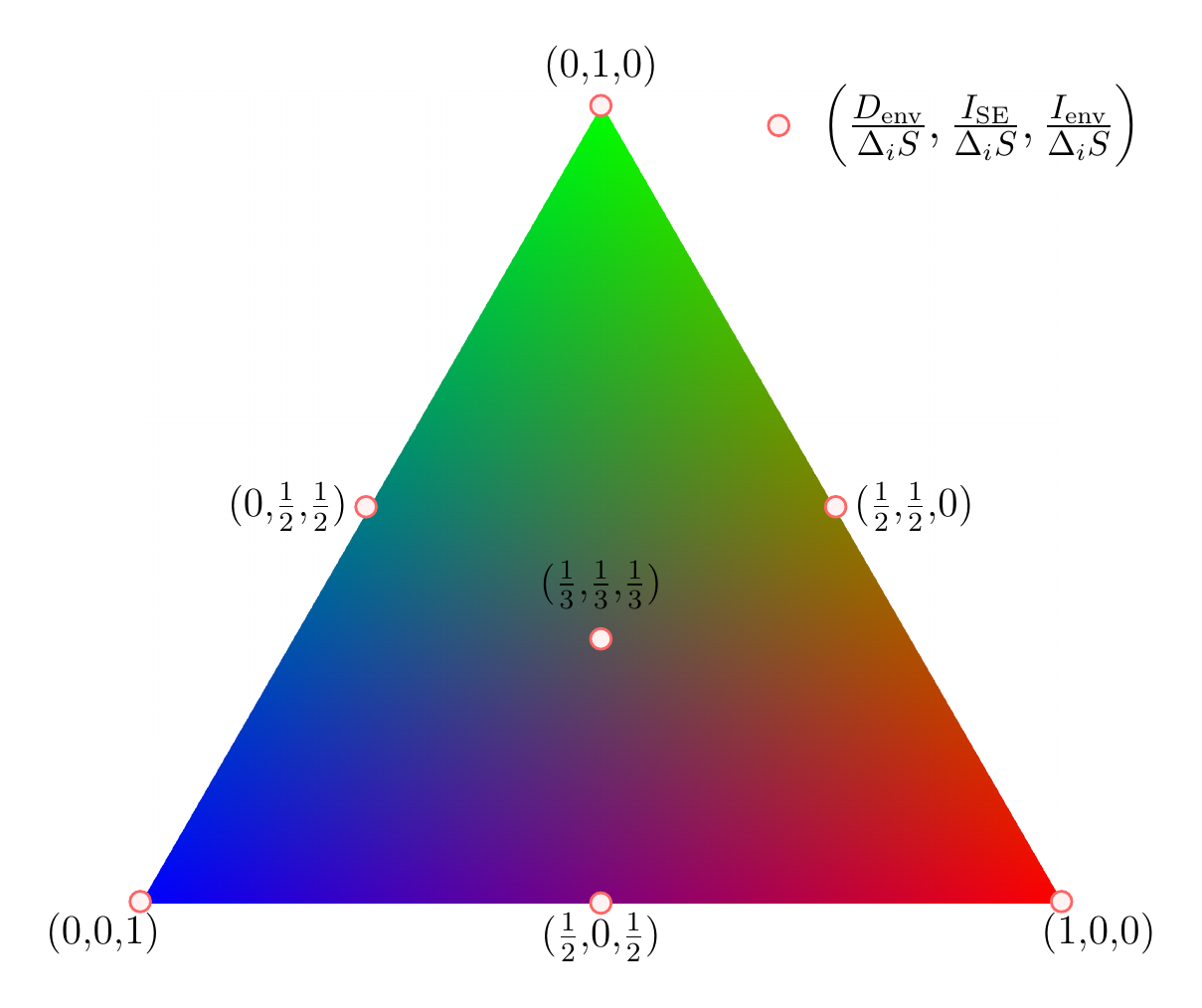}
\caption{Available colors. Full red indicates a very high percentage of the contribution coming from $D_{\mathrm{env}}$, full green from $I_{\mathrm{SE}}$, full blue from $I_{\mathrm{env}}$.  \label{fig:RGBmap}}
\end{figure}

From the numerical simulations, we find that the choice of parameters $\gamma$ and $T$ influences which quantity is most prominent. See for examples Figs. \ref{fig:contr_nd}a, \ref{fig:contr_nd}b and \ref{fig:contr_nd}c, which show the time dependent evolution of the three quantities, each at different parameter regimes; it is clear that $I_{\mathrm{env}}$ is not always the major contribution, but rather any quantity can be the winner for suitable values of $\gamma$ and $T$. 
To have a better grasp on which parameter ranges favour which quantity, we can look at the long time relaxation limit for many parameter pairs $(\gamma, T)$, and compute for each quantity the percentage of contribution to entropy production, namely the map:
\begin{equation}\label{RGBmap}
(\gamma,T)\longrightarrow \left.\left({D_{\mathrm{env}}(t) \over \Delta_iS(t)} , {I_{\mathrm{SE}}(t) \over \Delta_iS(t)} , {I_{\mathrm{env}}(t) \over \Delta_iS(t)}\right)\right|_{t=t_{\mathrm{max}}} \; .
\end{equation}
We take values of $\gamma$ going from ultra-weak coupling to strong coupling (0 to 2 in units of $\omega_0$) and of $T$ going from very low to high temperatures (0 to 5 in units of $\omega_0$ and $\mathrm{k_B}$). It is possible in this way to visually understand which areas of the plane $(\gamma, T)$ are comprised of which contribution. Specifically, one can take a colormap associating to each point $(\gamma, T)$ a RGB color defined by the three values of the map \eqref{RGBmap}. Since these three percentages have the requirement that their sum has to equal to 1, the possible colors are those found in a triangular shaped planar subset of the RGB color cube, namely the ones in Fig.~\ref{fig:RGBmap}. 

\begin{figure}[htp!]
  \includegraphics[clip,width=\columnwidth]{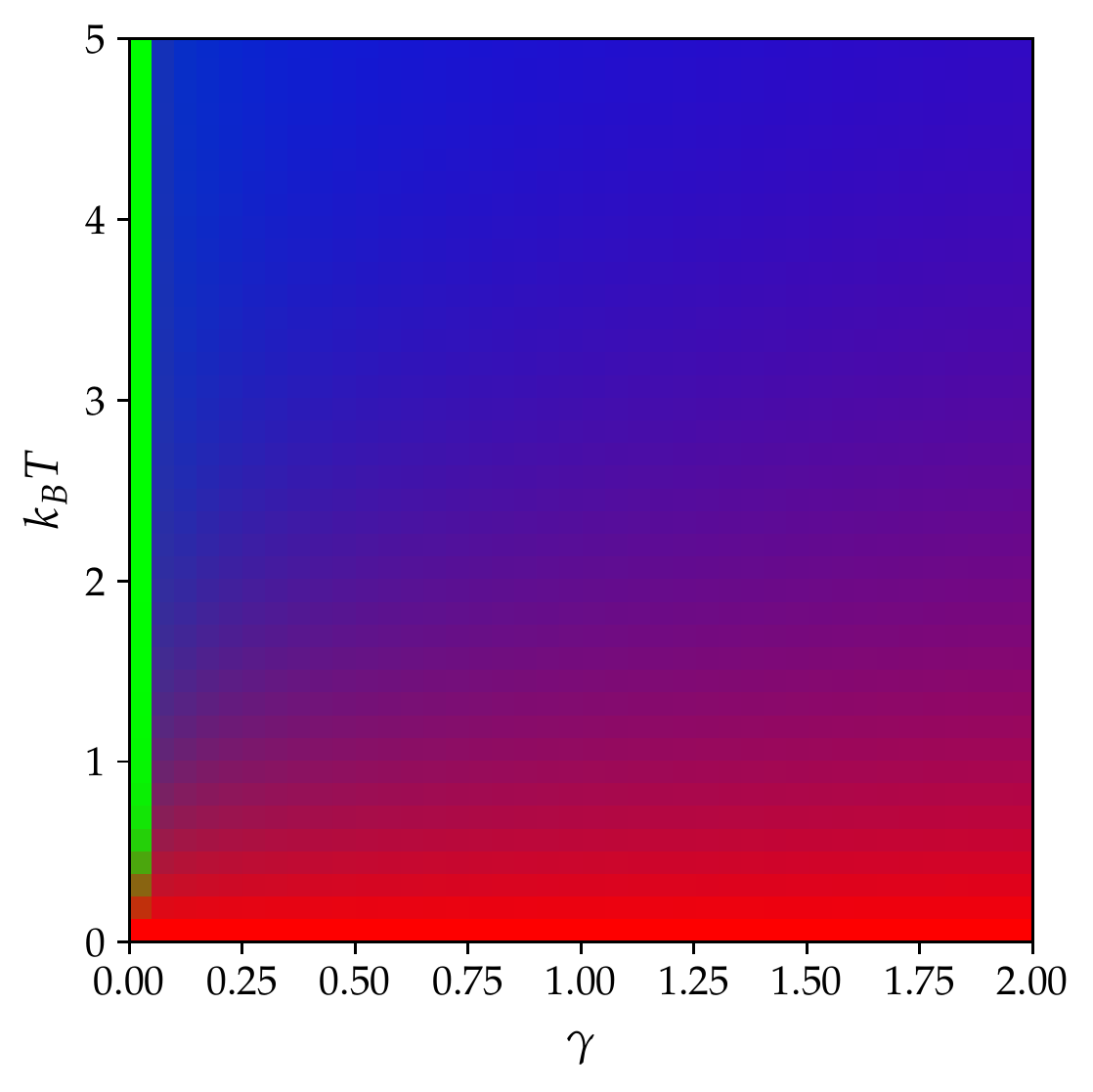}
\caption{Colormap of the composition of entropy production in absence of driving.}\label{fig:colormap_nd}
\end{figure}

The non-driven case is plotted this way in Figure \ref{fig:colormap_nd}, showing a clear red area at low temperatures, indicating a very strong percentage of $D_{\mathrm{env}}$ to the total entropy production.
For very weak coupling strength $\gamma$ (at not too low temperatures) the winning quantity is the mutual information $I_{\mathrm{SE}}$ with a border at around $\gamma=0.01$. This range corresponds to the limit in which the Caldeira-Leggett master equation in Lindblad form \cite{Caldeira1983} is applicable, namely weak-coupling and high temperature limit. The rest of the values, namely medium to strong coupling and medium to high temperature, are the reign of the intra-environment correlations $I_{\mathrm{env}}$, similarly to what was found in \cite{Esposito2019}. For a certain range of temperature ($\mathrm{k_B}T/\omega_0 \sim 1 - 1.5$) the purple hues indicate a mixing of $I_{\mathrm{env}}$ and $D_{\mathrm{env}}$ in comparable measure, see for more detail the time evolution in that range in Fig.~\ref{fig:contr_nd}d. 

\begin{figure}[htp!]
  \includegraphics[clip,width=\columnwidth]{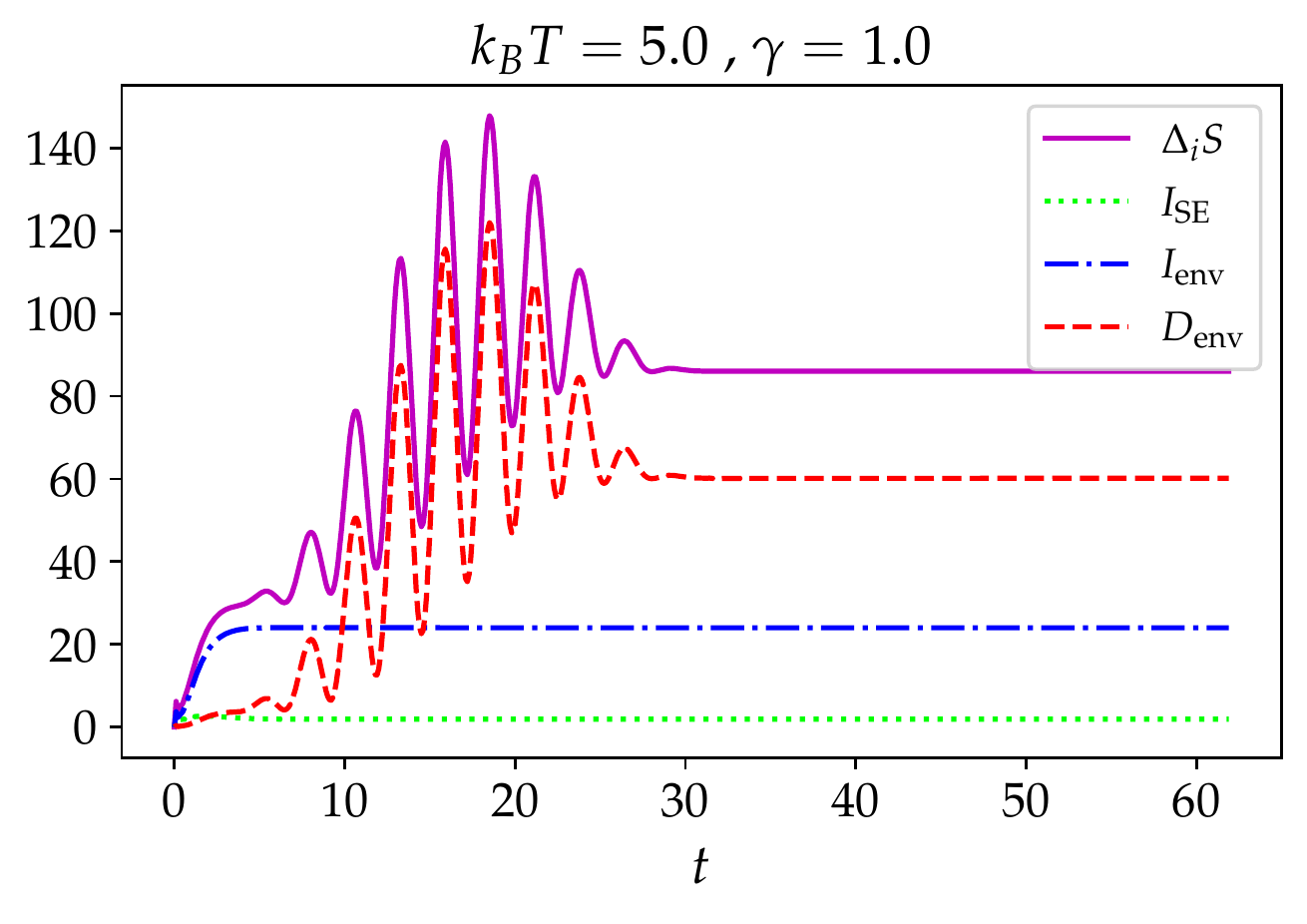}
\caption{Time evolution of the three contributions to entropy production $\Delta_iS^{\mathrm{ELB}}$ under strong driving ($F_0=10$), at high temperature and coupling ($\gamma = 1$, $\mathrm{k_B}T=5$). $D_{\mathrm{env}}$ is the dominant contribution.}\label{fig:contr_driving}
\end{figure}

\begin{figure}[htp!]
\includegraphics[clip,width=\columnwidth]{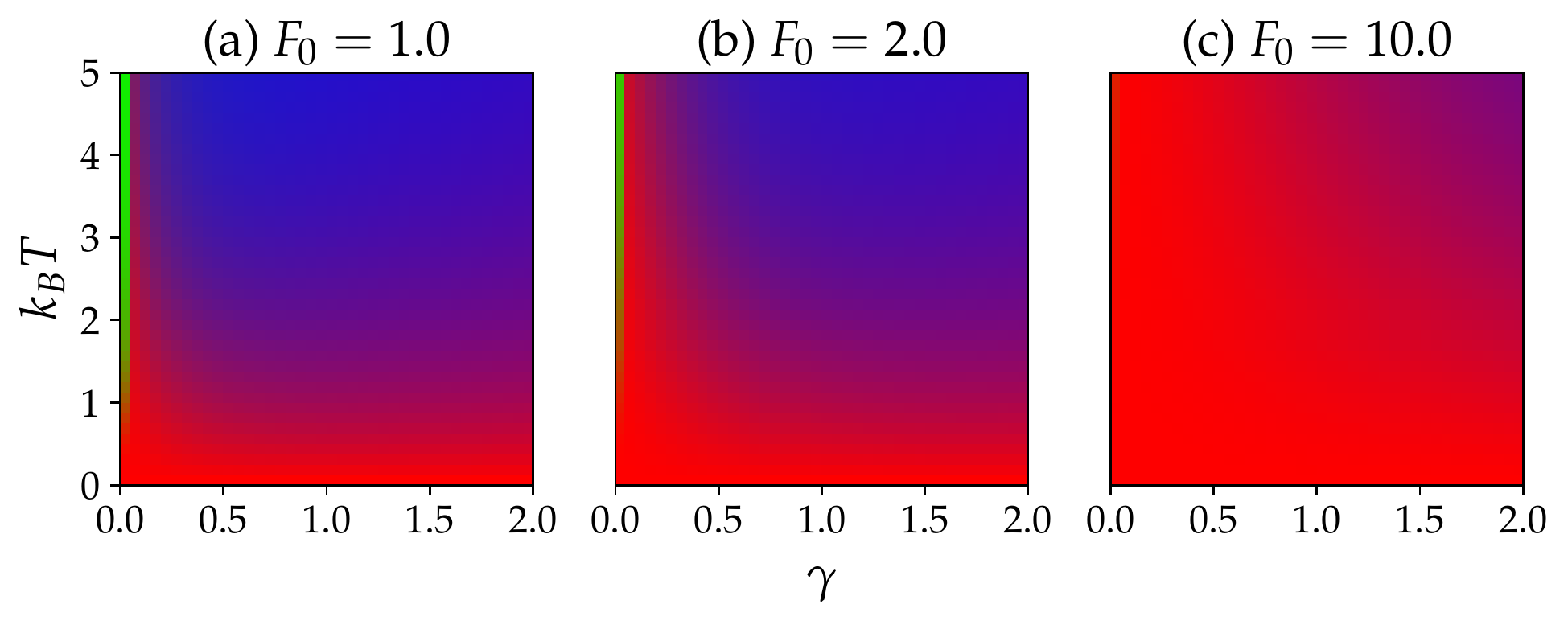}
\caption{Colormap of the composition of entropy production in presence of driving with various amplitudes, at an oscillating frequency of $\omega_f=1.2$.}\label{fig:colormap_F}
\end{figure}

\begin{figure}[htp!]
  \includegraphics[clip,width=\columnwidth]{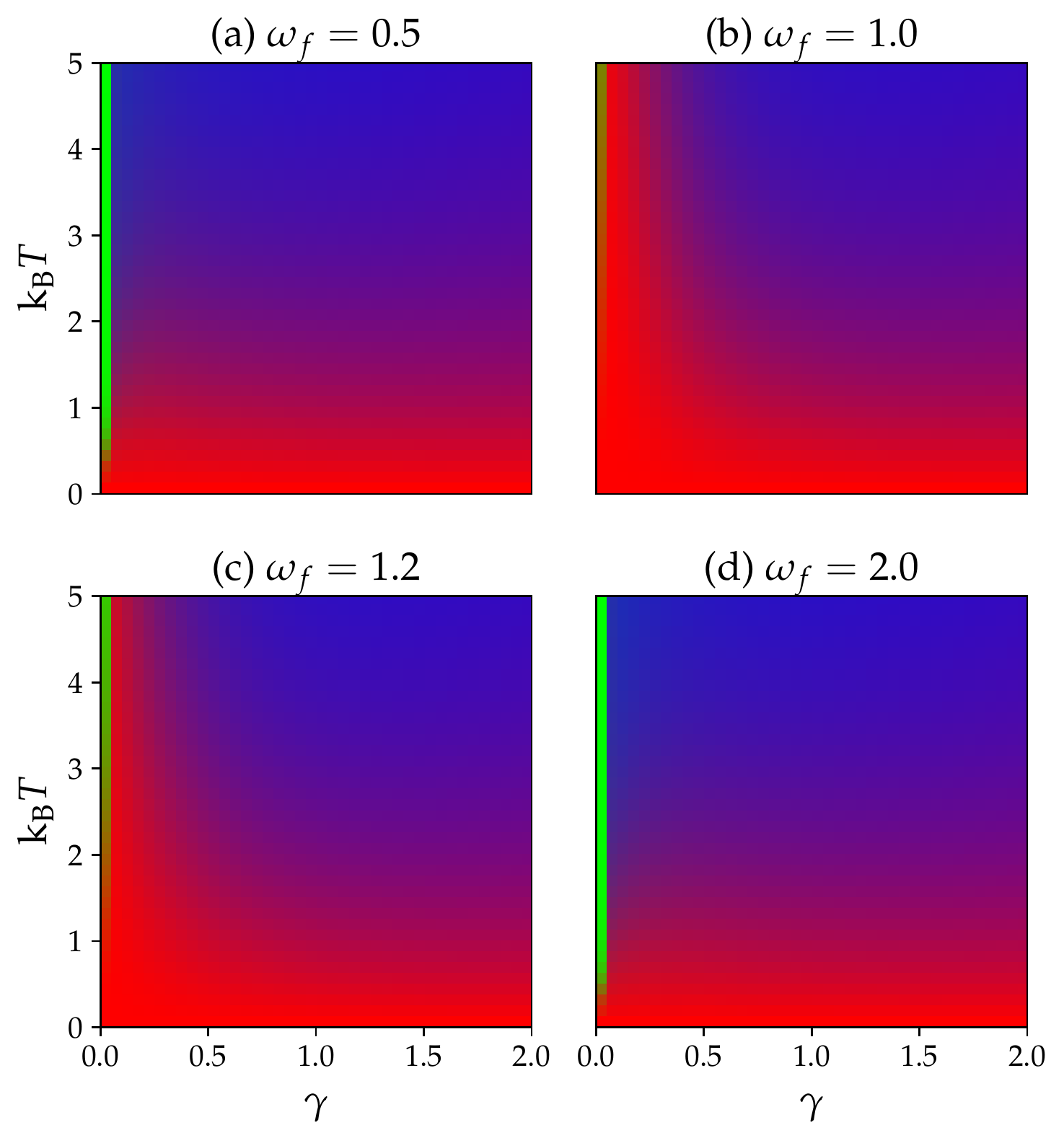}
\caption{Colormap of the composition of entropy production in presence of driving at various frequencies, with driving amplitude of $F_0=2$.}\label{fig:colormap_freq}
\end{figure}

Driving the central oscillator changes the composition of the entropy production; as already discussed, the quantity $D_{\mathrm{env}}$ is favoured, as it is the only one which is affected by driving, while $I_{\mathrm{SE}}$ and $I_{\mathrm{env}}$ remain unaltered. This is evident, for example, from the time evolution of the components for high coupling and temperature when strong driving is added, Fig. \ref{fig:contr_driving}: $D_{\mathrm{env}}$ clearly dominates, while in absence of driving the determining contribution had been $I_{\mathrm{env}}$. 
In terms of color areas for the long time limit, this translates into an expansion of the red area (dominance of $D_{\mathrm{env}}$) into a broader range of coupling and temperature. This expansion increases for increasing driving amplitude $F_0$, as can be seen from the plots in Fig.~\ref{fig:colormap_F}. 
The driving frequency $\omega_f$ also influences this behaviour, showing a larger stretch of the red area, especially in the lower range of $\gamma$, for frequencies near resonance ($\omega_f \sim \omega_0$), see Fig.~\ref{fig:colormap_freq}.

\subsection{System-bath entanglement}\label{sec:res_entanglement}

Working with Gaussian states enables us to also investigate another aspect of correlations in the Caldeira-Leggett model, namely whether quantum correlations in the form of entanglement are formed between the central oscillator and the heat bath. To do this, we examine the trend in time of the logarithmic negativity \eqref{logneg}, which uniquely shows whether entanglement is present or not (see Sec.~\ref{sec:entanglement}), once again for different ranges of the parameters $\gamma$ and $T$. We are moreover interested in comparing this quantity with the evolution of the mutual information $I_{\mathrm{SE}}$ between the system and the environment, which represents the total classical and quantum correlations. We have previously argued that it is sufficient to restrict this analysis to the undriven case $F_0=0$, as the displacement effect of driving on the system does not affect the value of mutual information, nor that of logarithmic negativity.

\begin{figure}[htp!]
\includegraphics[clip,width=\columnwidth]{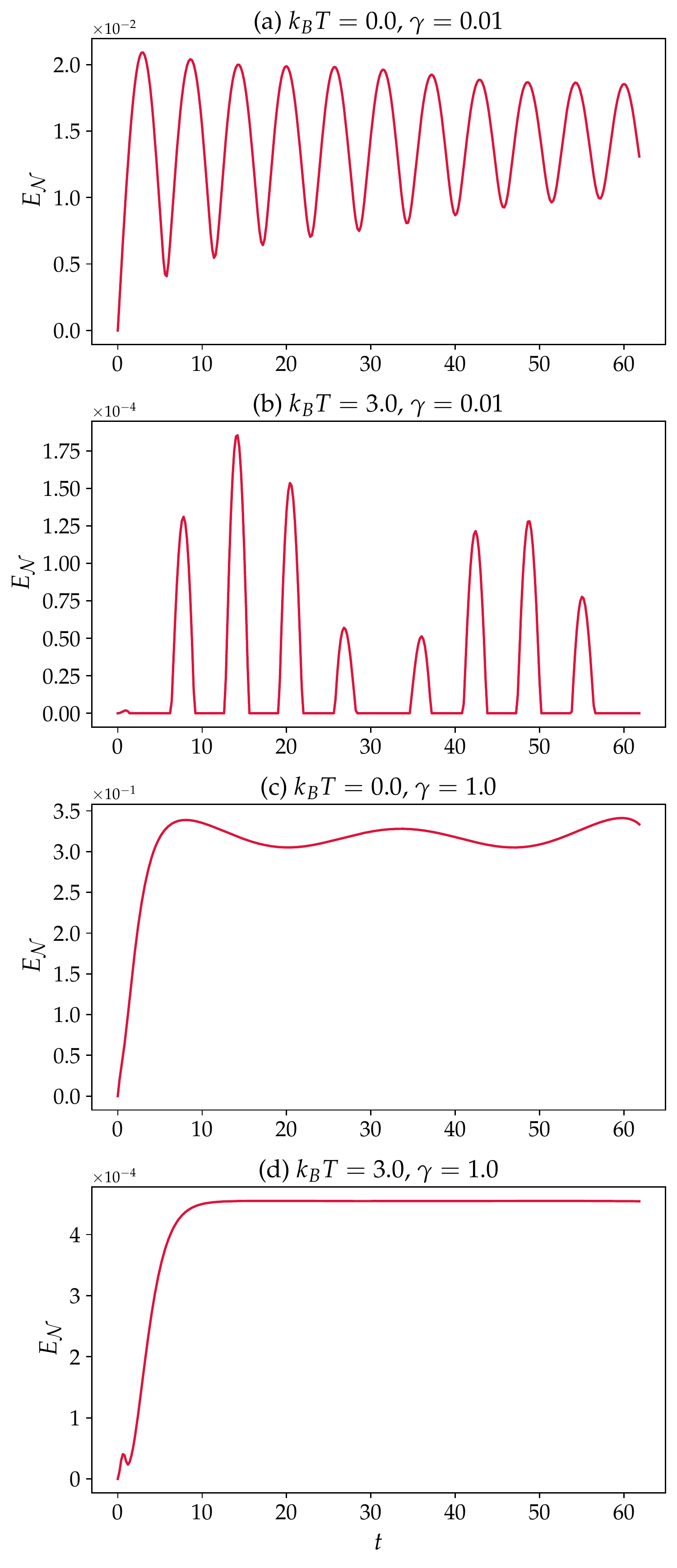}
\caption{System-bath entanglement measure for the Caldeira-Leggett model in absence of driving for four different ranges of coupling and temperature.}\label{fig:ent_cases}
\end{figure}

\begin{figure}[htp!]
  \includegraphics[clip,width=\columnwidth]{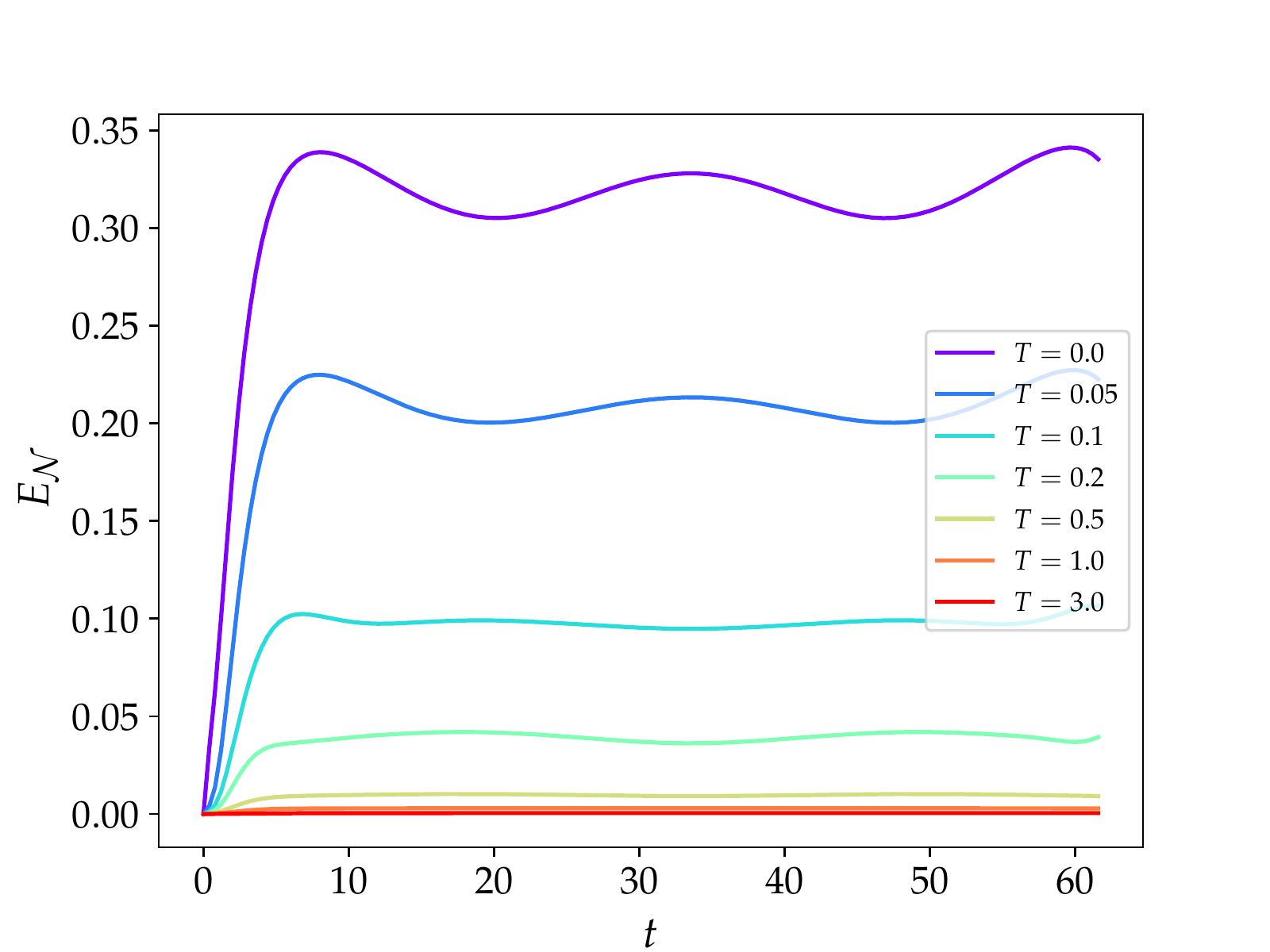}
\caption{Trend in time of logarithmic negativity at different temperatures, coupling $\gamma=1$.}\label{fig:entT}
\end{figure}

\begin{figure}[htp!]
  \includegraphics[clip,width=\columnwidth]{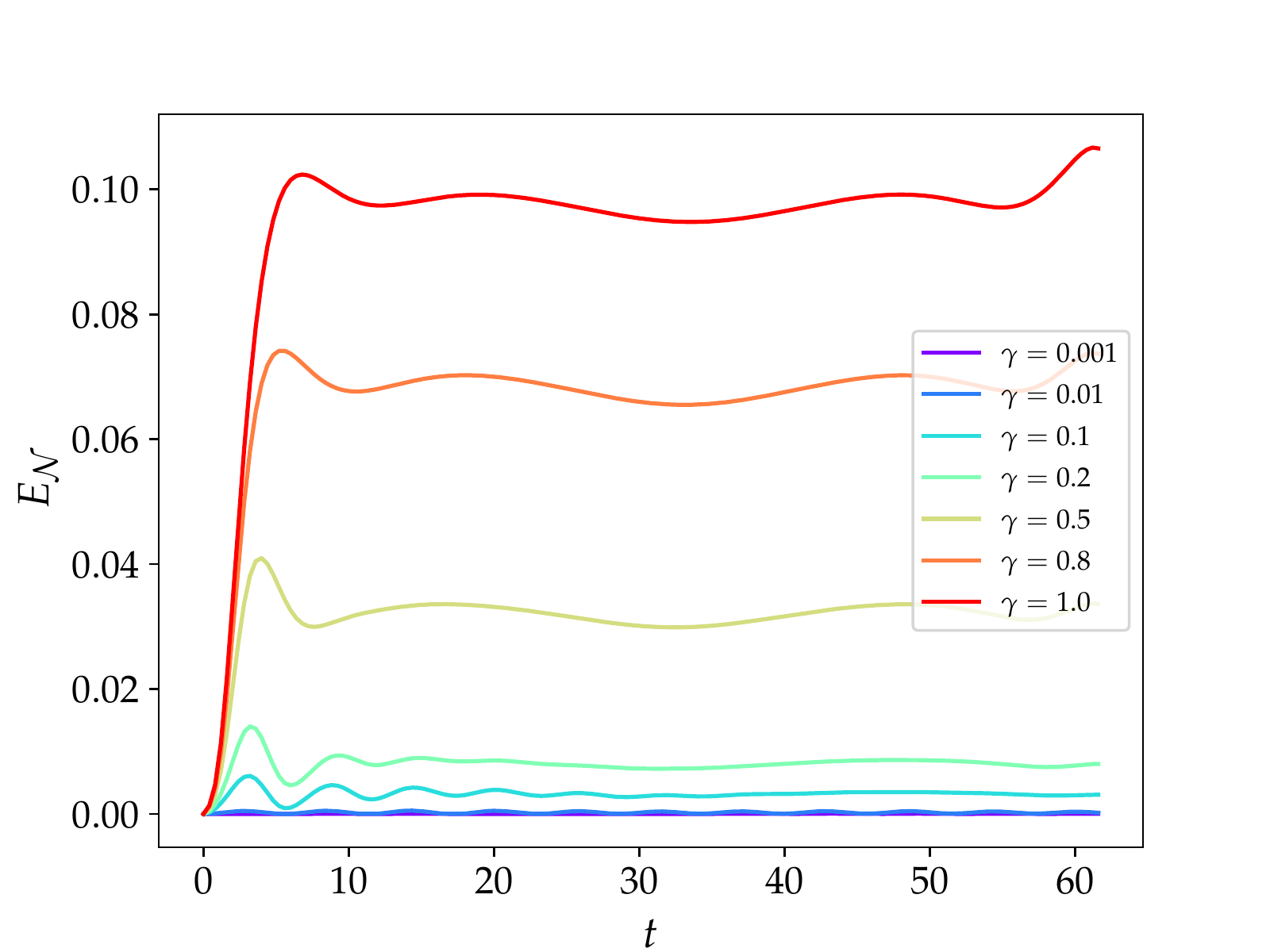}
\caption{Trend in time of logarithmic negativity at different coupling strength, at temperature $\mathrm{k_B}T=0.1$.}\label{fig:entg}
\end{figure}

First of all, from the simulations of logarithmic negativity we notice that there is entanglement generation between the system and environment (non-zero values of $E_{\mathcal{N}}$) for most parameter values, see Figs.~\ref{fig:ent_cases}a-\ref{fig:ent_cases}d. 
In general, the trend of entanglement looks for the most part oscillatory, with a frequency similar to $\omega_0$ (Fig.~\ref{fig:ent_cases}a, \ref{fig:ent_cases}b); for small coupling and higher temperatures, the logarithmic negativity even goes back down to zero periodically, exhibiting recurrent sudden death \cite{Eberly2009} and sudden birth of entanglement, showing that during the evolution in time the system and the bath can entangle and disentangle repeatedly (Fig.~\ref{fig:ent_cases}b). This oscillatory behaviour and recurrences seem to be typical for the evolution of entanglement in open quantum systems \cite{Aolita2015}. In the context of continuous variable systems, similar trends of bipartite entanglement have been predicted between two oscillators coupled to the same reservoir \cite{Paz2008,Horhammer2008} using the exact master equation for quantum Brownian motion (Hu, Paz and Zhang, \cite{Hu1992}). It is possible that the recurrences of entanglement, especially the more extreme ones we observe at lower values of $\gamma$ (Fig.~\ref{fig:ent_cases}a, \ref{fig:ent_cases}b), are a consequence of non-Markovian effects from the bath. Indeed, the oscillations seem to flatten down with increasing temperature (Fig.~\ref{fig:entT}) and appear more dramatic at intermediate values of coupling (Fig.~\ref{fig:entg}), which is precisely the regime of non-Markovianity for the Caldeira-Leggett model \cite{Einsiedler2020}.

\begin{figure}[htp!]
  \includegraphics[clip,width=\columnwidth]{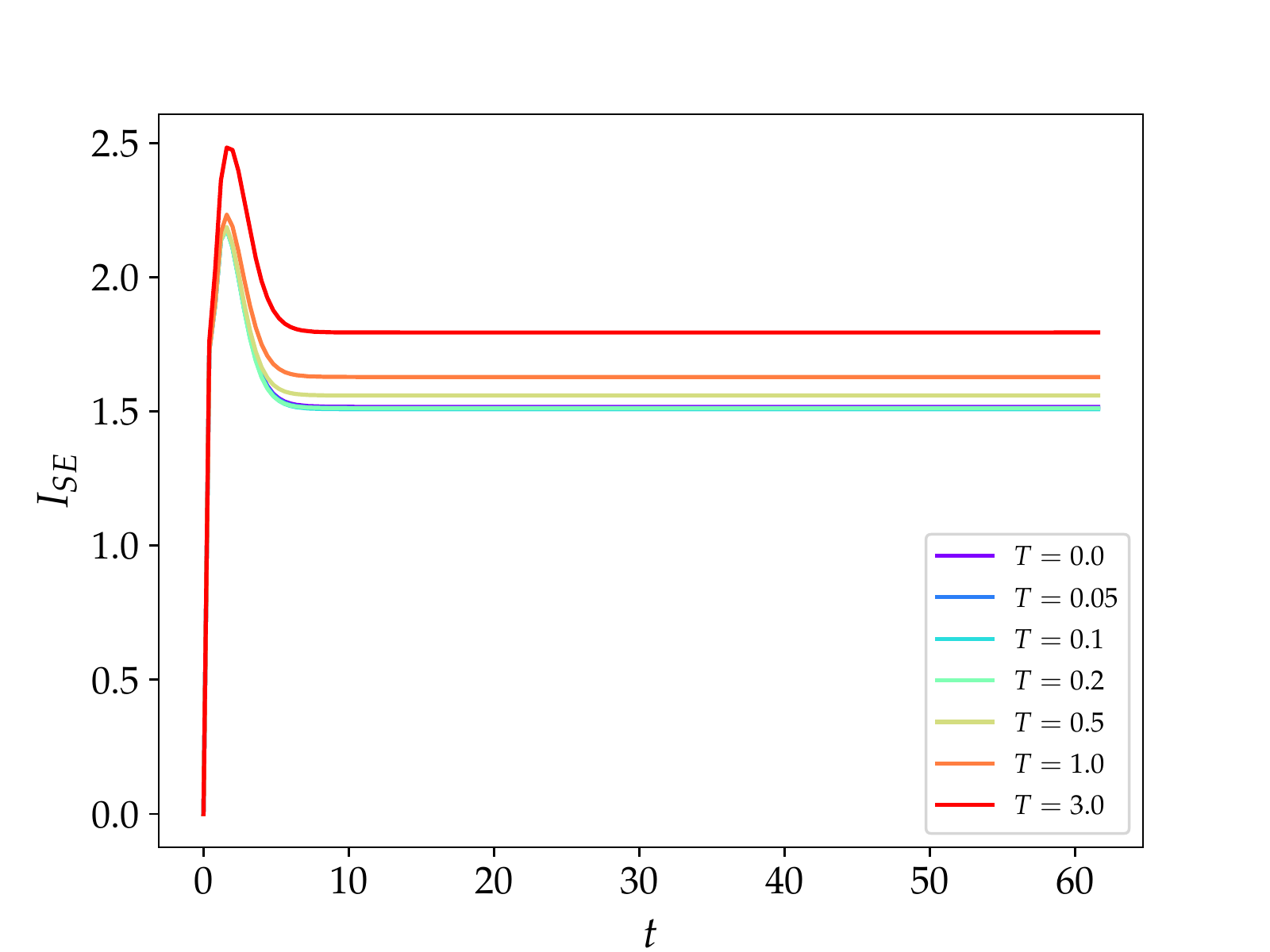}
\caption{Trend in time of mutual information at different temperatures, coupling $\gamma=1$.}\label{fig:iseT}
\end{figure}

\begin{figure}[htp!]
  \includegraphics[clip,width=\columnwidth]{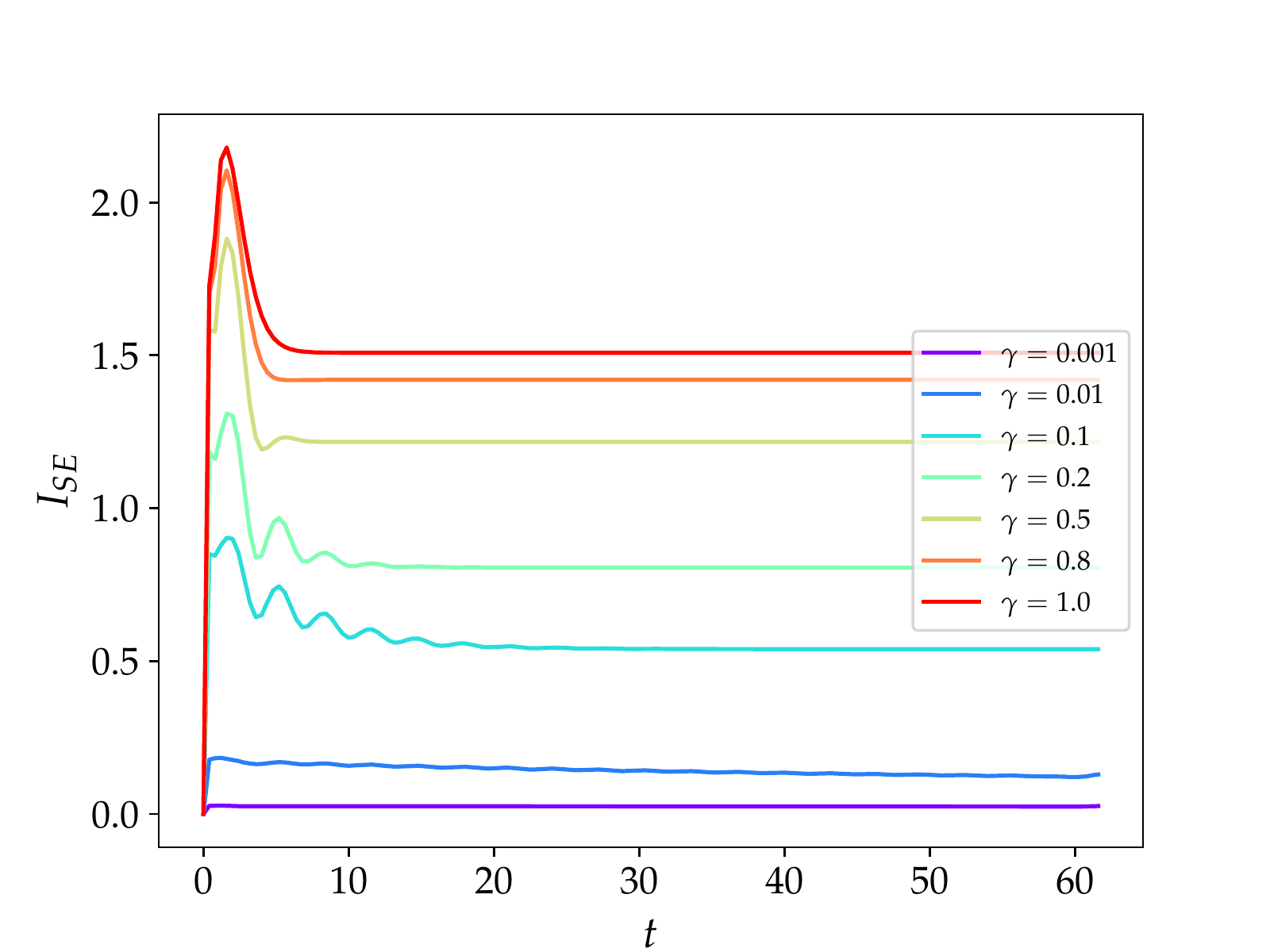}
\caption{Trend in time of mutual information at different coupling strength, at temperature $\mathrm{k_B}T=0.1$.}\label{fig:iseg}
\end{figure}

On another note, comparing the trend of $E_{\mathcal{N}}$ for different values of temperature, one can see that the amount of entanglement produced is suppressed at higher temperatures (Fig.~\ref{fig:entT}), with a more drastic variation in the range $T\sim 0.05-0.5$. On the contrary, performing the same comparison for mutual information, Fig.~\ref{fig:iseT}, reveals that the amount of total correlations increases with increasing temperature.
This indicates that raising the temperature leads to an increase in the correlations between the system and the reservoir which is surely not due to entanglement. Carrying out the simulations for different values of the coupling strength $\gamma$ shows (Fig.~\ref{fig:entg}) that the higher is the coupling, the more entanglement is formed.
This holds true also for the total correlations, as raising the coupling increases mutual information (Fig.~\ref{fig:iseg}). Therefore, no conclusions can be really drawn as to how changing the coupling strength affects the influence of entanglement on the total correlations.

\section{Conclusions}\label{sec:conclu}

The exact solvability of the Caldeira-Leggett model for quantum Brownian motion under the assumption of an initial Gaussian state enabled us to perform a well rounded study of exact, yet non-trivial quantities and effects. In particular, the restriction to a finite number of bath modes permits the evaluation of thermodynamic quantities and of measures for correlations within different entities in the model, all of which are part of relevant recent discussion. The three main parts of our study are all concerned with different aspects, that are nonetheless related to each other.  To summarize:

In part \ref{sec:res_proposals} we compared two different definitions for entropy production, namely $\Delta_i S^{\mathrm{DL}}$ (only valid in the weak coupling regime) and $\Delta_i S^{\mathrm{ELB}}$. The non-driven case has already been studied \cite{Pucci2013}, and we reproduced consistent results. What is novel here is the addition of a driving force acting on the central oscillator; our results show, interestingly, that the two definitions no longer properly converge in the weak coupling and high temperature limit, pointing at possible fundamental discrepancies between the two approaches.

In part \ref{sec:res_contributions} we studied the three different contributions of $\Delta_i S^{\mathrm{ELB}}$, that were already subject of a recent study concerning a different model \cite{Esposito2019}, in which the surprising importance of intra-environment correlations has been shown. Our study sheds more light on the topic, showing how other contributions could steal the spotlight, depending on the range of certain parameters such as coupling and temperature. Furthermore, we saw that the addition of the driving force once again tweaks the results, strongly favouring the contribution coming from the distance of the single bath modes from their initial state. 

In part \ref{sec:res_entanglement} we exploited the PPT criterion as a necessary and sufficient condition for separability between the 1-mode Gaussian state of the system and the N-mode Gaussian state of the environment to study the presence of quantum correlations between system and bath. We found that there is indeed entanglement generation, which is more prominent for strong coupling and low temperatures; on the contrary, mutual information increases with temperature, implying that raising temperature leads to the appearance of additional correlations which cannot be due to entanglement. Moreover, we witness a predominantly oscillatory behaviour of entanglement with time, in conjuction with sudden death and rebirth of entanglement for particular parameter choices. The analogy of this behaviour with the one of entanglement which is produced within composite systems coupled to non-Markovian reservoirs suggests that there is more to explore about the effects of non-Markovianity on entanglement generation, and about the role that the system-bath correlations may play on the ones within the system. 

In conclusion, our three different studies on entropy production and correlations, which all have the advantage of relying on exact quantities, add interesting information to the recent developments in the topics of quantum themodynamics and quantum correlations. As a potential main insight surging from this work, we believe that the inclusion of driving in model systems should be paramount in future investigations that aim at a better formulation of quantum thermodynamics. 

\begin{acknowledgments}
This project has received funding from the European Union’s Framework Programme for Research and Innovation Horizon 2020 (2014-2020) under the Marie Skłodowska-Curie Grant Agreement No. 847471.
\end{acknowledgments}

\bibliography{biblio}

\end{document}